\documentclass[11pt]{article}
\usepackage{epsfig} 
\newcommand{\sect}[1]{ \section{#1} \setcounter{equation}{0} } 

\newcommand{\Aslash}{A \! \! \! /} 
\newcommand{\kslash}{k \! \! \! /} 
\newcommand{\pslash}{p \! \! \! /} 
\newcommand{\qslash}{q \! \! \! /} 
\newcommand{\xslash}{x \! \! \! /} 
\newcommand{\zslash}{z \! \! \! /} 
\newcommand{\partialslash}{\partial \! \! \! /} 
 
\newcommand{\al}{\alpha}
\newcommand{\Del}{\Delta}
\newcommand{\Nc}{N_{\!c}} 
\newcommand{\Nf}{N_{\!f}} 
\newcommand{\MSbar}{\overline{\mbox{MS}}} 
\def \Tr {\mbox{Tr\,}}
\def \tr {\mbox{tr}}

\begin{document}
\hfill LTH--469  

\hfill NTZ 27/1999 

\hfill TPR--99--21

{\begin{center}
{\LARGE {Computation of quark mass anomalous dimension at $O(1/\Nf^2)$ in 
quantum chromodynamics} } \\ [8mm]
{\large M. Ciuchini$^a$, S.\'{E}. Derkachov$^b$\footnote{e-mail: 
Sergey.Derkachov@itp.uni-leipzig.de}, J.A. 
Gracey$^a$\footnote{e-mail: jag@amtp.liv.ac.uk} \& A.N. 
Manashov$^c$\footnote{e-mail: manashov@heps.phys.spbu.ru} \\ [3mm] } 
\end{center} 
} 

\begin{itemize}
\item[$^a$] Theoretical Physics Division, Department of Mathematical Sciences, 
\\
University of Liverpool, Liverpool, L69 7ZF, United Kingdom. 
\item[$^b$] Institut f\"{u}r Theoretische Physik, Universit\"{a}t Leipzig,
\\
Augustusplatz 10, D-04109 Leipzig, Germany \& \\ 
Department of Mathematics, St Petersburg Technology Institute, \\
Sankt Petersburg, Russia  
\item[$^c$] Institut f\"ur Theoretische Physik, Universit\"at
Regensburg,\\ D-93040 Regensburg, Germany \& \\
Department of Theoretical Physics, State University of St Petersburg, \\ 
Sankt Petersburg, 198904 Russia.
\end{itemize}

\vspace{3cm} 
\noindent 
{\bf Abstract.} We present the formalism to calculate $d$-dimensional critical 
exponents in QCD in the large $\Nf$ expansion where $\Nf$ is the number of 
quark flavours. It relies in part on demonstrating that at the $d$-dimensional 
fixed point of QCD the critical theory is equivalent to a non-abelian version 
of the Thirring model. We describe the techniques used to compute critical two 
and three loop Feynman diagrams and as an application determine the quark wave 
function, $\eta$, and mass renormalization critical exponents at $O(1/\Nf^2)$ 
in $d$-dimensions. Their values when expressed in relation to four dimensional 
perturbation theory are in exact agreement with the known four loop $\MSbar$ 
results. Moreover, new coefficients in these renormalization group functions 
are determined to six loops and $O(1/\Nf^2)$. The computation of the exponents 
in the Schwinger Dyson approach is also provided and an expression for $\eta$ 
in arbitrary covariant gauge is given.  

\newpage 

\sect{Introduction.} 

Recent developments in the area of multiloop calculations in quantum 
chromodynamics, (QCD), have included the determination of the $\MSbar$ four
loop $\beta$-function, \cite{RVL97}, and the four loop quark mass anomalous 
dimension, \cite{VLR97,C97}. These calculations involved current state of the
art analytic and symbolic manipulation techniques in the evaluation of the 
order of $10,000$ Feynman diagrams. Such high order calculation are necessary
to refine our understanding, for example, of the running of the quark masses,
\cite{VLR97,C97}. Indeed such four loop calculations have built on the early 
lower order work of {\em several} decades of \cite{GW73,CJ74,ET79,TVZ80,LV93} 
for the QCD $\beta$-function and \cite{NRT79,TNW81,T82} for the quark mass
anomalous dimension. Whilst we believe the results of \cite{RVL97} and 
\cite{VLR97,C97} will not be superseded by a five loop calculation for quite
some time, it is possible to probe the higher order structure of the QCD 
renormalization group, (RG), functions using techniques alternative to 
conventional perturbation theory. One such approach is the large $\Nf$ 
expansion where $\Nf$ is the number of quark flavours. In this method the 
Feynman diagrams contributing to the determination of an RG function are 
reordered according to the number of quark loops and evaluated by considering
at leading order those graphs which are simply a chain of bubbles. The next 
order is represented by those bubble chains which have either one vertex or
self energy correction and so on. Clearly this large $\Nf$ reordering will 
cover the information already contained in explicit perturbative loop 
calculations but will reveal new information beyond the current orders. For 
simple scalar field theories such as the $O(N)$ $\phi^4$ theory or the $O(N)$
nonlinear $\sigma$ model, the large $N$ technique has been developed to 
determine information on the RG functions at $O(1/N^3)$, \cite{VPH,VPH82}. 
That impressive programme of Vasil'ev et al exploited ideas from critical 
renormalization group theory and determined rather than the RG functions
themselves, the related critical exponents. These correspond to the RG 
functions evaluated at the $d$-dimensional fixed point of the $\beta$-function
which in $O(N)$ $\phi^4$ theory is the Wilson-Fisher fixed point. The 
techniques developed in the early work on scalar theories, 
\cite{VPH,VPH82,VN82}, have been applied to QCD at leading order in $1/\Nf$ in
\cite{JG93,JG96}. Further, information on the anomalous dimension of the 
twist-$2$ operators fundamental to the operator product expansion used in deep 
inelastic scattering has been produced as a function of the operator moment, 
\cite{JGJB}. Although it is important that these calculations are extended to 
next order in $1/\Nf$, it turns out that there are several issues which need to
be addressed. First, no rigorous proof exists of the critical equivalence of 
QCD and the so called nonabelian Thirring model, (NATM), which underlies all 
the large $\Nf$ computations. The original observation of the connection 
between the models in \cite{H2} was effectively at $O(1/\Nf)$. Second, the 
analytic regularization commonly used in $1/N$ calculations breaks the gauge 
invariance of the theory. As far as we are aware all other regularizations 
spoil the masslessness of the propagators and this therefore makes higher order
calculations virtually impossible. We note that dimensional regularization, 
which is ordinarily used in conventional perturbation theory, is not applicable
in large $\Nf$ work since the theory remains divergent in arbitrary dimensions.
This second obstacle is much more serious and might seem to destroy the 
possibility of developing a sensible $1/\Nf$ scheme. The resolution of the 
issues of the proof of the critical equivalence of QCD and the NATM as well as 
demonstrating that the critical exponents evaluated in the latter model using 
non-gauge invariant regularizations do in fact match those of QCD are of 
extreme importance and represent the central results of this paper. By way of 
application and in order to support the scheme which will be developed in 
practical calculations we examine the quark and mass anomalous dimensions and 
determine them at new order in the large $\Nf$ expansion which is $O(1/\Nf^2)$.
A preliminary version of our results was given in \cite{CDGM}.

There are various motivations for examining these two RG functions. First, as
in conventional perturbation theory one always needs to deduce the wave 
function renormalization of the fundamental fields of a theory before 
considering the renormalization of the other parameters and operators in a 
Lagrangian. Likewise in the large $\Nf$ approach the evaluation of the wave
function exponent, $\eta$, needs to be determined first. Indeed in the context 
of the phenomenological application of the large $\Nf$ technique the programme 
of evaluating anomalous dimensions of the twist-$2$ operators of deep inelastic
scattering at $O(1/\Nf^2)$ cannot proceed prior to the determination of $\eta$ 
at the same order. Second, although $\eta$ is fundamental it depends on the 
covariant gauge parameter introduced in the gauge fixing of the NATM 
Lagrangian. Consequently, it is not as fundamental a quantity as a gauge 
independent exponent. Therefore, to further understand the large $\Nf$ method 
in QCD at $O(1/\Nf^2)$ we have undertaken to consider the gauge independent 
quark mass anomalous dimension. Indeed as part of our calculations we address 
the issue of the choice of gauge in the large $\Nf$ method. Whilst it may seem 
that the evaluation of two exponents at $O(1/\Nf^2)$ is a formidable task, it 
turns out that the computation of the quark mass dimension does not represent a
significant amount of additional calculation as there is, in fact, a close 
relation between the fermionic graphs for the wave function and mass dimension 
which we will establish. 

It is worth noting that our detailed calculation relies on the novel method 
of~\cite{DM,DMrus98} for the computation of critical exponents in $1/N$ 
expansion. In comparison with the standard technique~\cite{VPH} based on the 
direct solution of the Schwinger-Dyson (SD) equations the latter approach
allows us to express the critical exponents through the corresponding 
renormalization constants whose explicit form is very similar to those computed
using dimensional regularization. The immediate advantage is that it allows one 
to use techniques similar to the infra-red rearrangement of conventional high 
order perturbative calculations to evaluate the three loop critical point 
Feynman diagrams needed for the present calculation. This approach, which was 
originally developed to analyse the $O(N)$ nonlinear $\sigma$ model appears to 
be universal, and can be easily extended to the case we will consider. Together
with some calculational shortcuts it makes the evaluation of almost all the 
relevant graphs quite straightforward and simplifies computations 
significantly. In addition we will also present the Schwinger-Dyson formalism 
for the determination of $\eta$ at $O(1/\Nf^2)$ in QCD. This builds naturally 
on the earlier large $\Nf$ calculations in quantum electrodynamics, (QED), 
\cite{JG91,JG94,JG93m}. In those papers the original technique of \cite{VPH} in
solving the SD equations was followed and in extending the formalism to QCD 
here will provide an interested reader with a unique opportunity to compare 
both approaches.  Finally, it is worth stressing that the consistency of our 
final results for the quark and mass anomalous dimensions at $O(1/\Nf^2)$ with 
the corresponding four loop perturbative results of 
\cite{NRT79,TNW81,T82,VLR97,C97} provides a nontrivial check on the validity of
either approach.
 
The paper is organised as follows. Section 2 is devoted to introducing 
background formalism and providing the proof of the critical equivalence of QCD
and NATM. In section 3 we develop the large $\Nf$ method explicitly in the 
context of QCD and the NATM. The details of the evaluation of the three loop 
diagrams to determine the quark anomalous dimension are described at length in 
section 4. Section 5 is devoted to the development of the Schwinger Dyson 
formalism to determine the quark anomalous dimension. The computation of the 
extra three loop graphs needed to determine the mass anomalous dimension is 
given in section 6. The final results for the quark anomalous dimension and 
mass anomalous dimension are collected together in section 7 where we also 
derive new information on the coefficients of the respective RG functions. 
Several appendices contain results which were fundamental to our calculations.  

\sect{Background.}
The QCD Lagrangian in $d$ $=$ $4$ $-$ $2\epsilon$ dimensional Euclidean space 
reads
\begin{equation}
\label{LQCD}
{\cal L} ~=~ \bar\psi^{iI}\not\!\!D \psi^{iI} ~+~ \frac{1}{4g^2} 
F^a_{\mu\nu}F^{a \, \mu\nu} ~+~ \frac{1}{2\xi g^2}(\partial\cdot A)^2 ~+~ 
\partial_\mu \bar c^a \left(D^{\mu}c\right)^a ~,
\end{equation}
where $\psi^{iI}$ is the quark field belonging to the fundamental representation
of the colour group, $1$~$\leq$~$I$~$\leq$~$\Nf$, $A_\mu^a$ is the gluon field,
$c^a$ and ${\bar c}^a$ are the ghost fields in the adjoint representation of 
the colour group, $\xi$ is the covariant gauge parameter and $g$ is the 
coupling constant. The field strength tensor $F^a_{\mu\nu}$ and the covariant 
derivative are defined by 
\[
F_{\mu\nu}^a ~=~ \partial_\mu A_\nu^a ~-~ \partial_\nu A_\mu^a ~+~ f^{abc} 
A^b_\mu A^c_\nu
\]
and 
\[
D_\mu ~=~ (\partial_\mu-i A_\mu^a T^a) ~,
\]
where $T^a$ are the colour group generators in the corresponding representation
and $f^{abc}$ are the structure constants with $[T^a,T^b]$ $=$ $i f^{abc} T^c$.
To ensure the coupling constant, $g$, is dimensionless below four dimensions we
rescale it in the standard way by setting $g$ $\to$ $ M^{\epsilon}g$, where the
parameter $M$ has dimensions of mass. The partition functions of QCD are 
defined as
\begin{equation}
\label{corr}
\langle O_1(x_1)\ldots O_n(x_n)\rangle ~=~
Z^{-1}\int D\Phi\ O_1(x_1)\ldots O_n(x_n) \exp{\{-S\}} ~,
\end{equation}
where $\Phi\equiv\{A,\bar\psi,\psi,\bar c, c\}$ is the set of fundamental 
fields, $O_i(x_i)$ represent a basic field or a composite operator and $Z$ is a
normalizing factor. As usual, the divergences arising in the calculation 
of~(\ref{corr}) are removed by the renormalization procedure at each order of 
perturbation theory. Namely, provided that the averaging in~(\ref{corr}) is 
carried out with the renormalized action all correlators of the fields taken at
different space-time points will be finite. The renormalized action 
$S_R(\Phi,e)$, ($e\equiv\{g,\xi\}$), has the form of the action~(\ref{LQCD}) 
with the fields and parameters being replaced by the bare ones
\[
S_R(\Phi,e) ~=~ S(\Phi_0,e_0) ~, \ \ \ \ \Phi_0 ~=~ Z_\Phi\>\Phi ~,\ \ \ 
e_0 ~=~ Z_e\>e ~.
\]
Here $Z_\Phi$ $=$ $\{Z_A,Z_\psi,Z_c\}$ and $Z_e$ $=$ $\{Z_g,Z_\xi\}$ are the 
renormalization constants. However this procedure does not guarantee the 
finiteness of the Green functions with operator insertions, which contain 
additional divergences. To remove these extra divergences one should 
renormalize the composite operators as well. In general, the renormalized 
operator reads 
\begin{equation}
\left [ O_i \right]_R ~=~ \sum_k Z_{ik}O_k ~,
\end{equation}
where the operators $O_k$ have canonical dimensions equal to or less than those
of the original operator $O_i$, and $Z_{ik}$ is the mixing matrix of 
renormalization constants. An operator is called multiplicatively renormalized 
if the matrix $Z_{ik}$ is diagonal, giving $[O_i]_R$ $=$ $Z_i O_i$ where there 
is no summation over $i$.

The renormalization group equation (RGE) for the one-particle irreducible
$n$-point Green function with the insertion of $k$ multiplicatively 
renormalizable composite operators reads 
\begin{equation}
\label{RG}
\left( M\partial_M+\beta_g\partial_g+\beta_\xi \partial_\xi
-n_\Phi \gamma_\Phi+\sum_{i=1}^k \gamma_{O_i} \right) \, 
\Gamma(x_1,\ldots,x_{n+k},M,g,\xi) ~=~ 0 
\end{equation}
where we use the shorthand notation $n_\Phi\gamma_\Phi$ $=$ $n_A\gamma_A$ $+$ 
$n_\psi \gamma_\psi$ $+$ $n_c \gamma_c$. The RGE functions $\beta_g$, 
$\beta_\xi$, $\gamma_\Phi$, $\gamma_{O_i}$ which are the respective beta 
functions and the anomalous dimensions of the fields and composite operators, 
are defined as follows
\begin{equation}
\label{RGf}
\beta_g ~=~ M\frac{d}{dM}g(M) ~,\ \ \ \ \beta_\xi ~=~ M\frac{d}{dM}\xi(M) ~,
\ \ \ \ \gamma_{\Phi} ~=~ M\frac{d}{dM}\ln Z_{\Phi} ~,
\ \ \ \ \gamma_{O_i} ~=~ M\frac{d}{dM}\ln Z_{O_i} ~.
\end{equation} 
Formally, the parameters $g$ and $\xi$ enter (\ref{RG}) on the same footing 
and should be considered as independent charges. However, the gauge fixing 
paremeter $\xi$ is introduced into the theory by hand and cannot enter an 
expression for a physical quantity. Thus for proper (gauge invariant) objects
the term $\beta_\xi \partial_\xi$ drops out of (\ref{RG}), which then
takes the form of an RGE of a single charge theory.

Our subsequent analysis relies heavily on the existence of a non-trivial 
infra-red, (IR), stable fixed point $g_{*}$ of the $d$-dimensional 
$\beta$-function, $\beta_g(g_{*})$ $=$ $0$, for large values of $\Nf$. The 
$\beta$-function has been calculated in $\MSbar$ using dimensional 
regularization up to $O(a^6)$ terms, where $a$ $=$ $(g/2\pi)^2$ $=$ 
$\alpha_s/\pi$, in \cite{GW73,CJ74,ET79,TVZ80,LV93}. We only record the first 
few terms here, while the full four loop result can be found in \cite{RVL97},
\begin{eqnarray}
\label{betag}
\beta_a(a) &=& (d-4)a ~+~ 
\left[\frac23 T_F \Nf-\frac{11}{6}C_A \right]a^2 ~+~
\left[\frac12 C_FT_F \Nf+\frac56 C_A T_F \Nf -\frac{17}{12}C_A^2 \right] a^3 
\nonumber \\ 
&& -~ \left[ \frac{11}{72} C_F T_F^2 \Nf^2 +\frac{79}{432} C_A T_F^2 \Nf^2+
\frac{1}{16} C_F^2 T_F \Nf -\frac{205}{288} C_FC_A T_F \Nf\right.
\nonumber \\
&& \left. ~~~~ -\frac{1415}{864} C_A^2 T_F \Nf +\frac{2857}{1728} C_A^3\right ] 
a^4 ~+~ O(a^5) ~,
\end{eqnarray}
from which it follows that
\begin{equation}
\label{gc}
a_{*} ~=~ \frac{3\epsilon}{T_F \Nf} ~+~ \frac{1}{4T_F^2 \Nf^2} \left( 
33C_A\epsilon-\left[ 27C_F+45C_A \right]\epsilon^2 + O(\epsilon^3) \right) ~+~ 
O\left( \frac{1}{\Nf^3} \right) ~.
\end{equation}
The Casimirs for a general classical Lie group are defined by 
\begin{equation}
\Tr \left(T^a T^b \right) ~=~ T_F \delta^{ab} ~, \ \ \ \ T^a T^a ~=~ C_F  I ~,
\ \ \ \ f^{acd} f^{bcd} ~=~ C_A \delta^{ab} ~.  
\end{equation}

It immediately follows from (\ref{RG}) that the Green functions of gauge 
invariant operators are scale invariant at the critical point $g_{*}$. In other
words 
\[
G(\lambda x_i) ~=~ \lambda^{D_i} G(x_i) ~,
\] 
where $D_i$ is the scaling dimension of the corresponding Green function. 
Moreover, due to the IR nature of the fixed point, this index determines the 
power of the leading term of the IR asymptotic behaviour of the Green functions
($p_i\to 0$) near the critical points and this is stable against the 
perturbation of the action by IR irrelevant operators \cite{Zinn}. On the 
other hand Green functions of gauge dependent objects, such as the propagators 
of the basic fields which will in general depend on $\xi$, are not invariant 
under scale transformations. Although one may restrict attention from the 
outset to gauge independent quantities, since they have physical meaning, it is
also possible and convenient to choose $\xi$ so that {\it all} Green functions 
are scale invariant. Evidently, this is equivalent to the condition 
$\beta_\xi(g_{*},\xi_{*})$ $=$ $0$. Since 
\begin{equation}
\beta_\xi (g, \xi) ~=~ -~ 2\xi(\epsilon+\gamma_A+\beta_g/g) ~,
\end{equation} 
one concludes that the equation $\beta_\xi(g_{*},\xi_{*})$~$=$~$0$ has two 
solutions. One is $\xi_{*}$ $=$ $0$ whilst the other is 
$\gamma_A(g_{*},\xi_{*})$ $=$ $-$ $\epsilon$. Bearing in mind that our main aim
is the development of the $1/\Nf$ expansion we choose the first solution, $\xi$
$=$ $0$, since the latter gives $\xi$ $\sim$ $\Nf$, which leads to problems in 
the construction of the $1/\Nf$ scheme. The origin of the above two solutions 
for $\xi$ becomes more transparent if one tries to write down the most general 
form of the gluon propagator satisfying the requirements of both scale and 
gauge invariance. Indeed, scale invariance yields
\begin{equation}
G_{\mu\nu}(p) ~=~ \frac{M^{2\epsilon}}{(p^2)^\alpha} \left(
A P_{\mu\nu}^\perp+B P_{\mu\nu}^\parallel\right) ~,
\end{equation}
where $P^{\perp}_{\mu\nu}$ and $P^\parallel_{\mu\nu}$ are the transverse and
longitudinal projectors, respectively, and $A$ and $B$ are constants. As is 
well known, \cite{Zinn}, radiative corrections do not contribute to the 
longitudinal part of gluon propagator. Hence, $G_{\mu\nu}^\parallel$ $=$
$\xi g^2 M^{2\epsilon} P_{\mu\nu}^\parallel p^{-2}$. This implies, that 
if $\alpha$~$\neq$~$1$ then $\xi$ must vanish, $\xi$ $=$ $0$. On the other 
hand for $\xi$ $\neq$ $0$ then one must have $\alpha$ $=$ $1$ which is easy to
check is equivalent to $\gamma_A$ $=$ $-$ $\epsilon$ corresponding to the 
canonical dimension of the field. Earlier work concerning the relation of 
scaling and conformal symmetry in the context of gauge theories has been given
in \cite{VPP1,VPP2}. We also note that in an abelian gauge theory the condition 
$\gamma_A$ $=$ $-$ $\epsilon$ follows directly from the Ward identities that 
implies the scale invariance of {\it all} Green functions at the critical point
$g$ $=$ $g_{*}$ in any gauge \cite{JG93m}.

As is well known from the theory of the critical phenomena \cite{Zinn} physical
systems which look quite different at the microscopic level may exhibit the 
same behaviour at the phase transition point. An example of this universality 
is the fixed point relation between the Heisenberg ferromagnet and
$d$-dimensional $\phi^4$ field theory. In what follows we construct the theory
belonging to the same universality class as large $\Nf$ QCD but which has a
simpler structure. To this end we develop the $1/\Nf$ expansion correlators of 
the type given in~(\ref{corr}) and analyse their behaviour in the IR region. As
usual, the first step is to integrate over the fermion fields in the functional
integral to obtain the effective action for the gluon field
\begin{equation}
\label{Seff}
S^{\mbox{eff}}_{A} ~\equiv~ \Nf\left (- \, 
\tr\ln(\!\not\!\partial-i\!\not\!\!A^{a}T^{a})
+\frac{M^{-2\epsilon}}{4\bar{g}^2} (F^a_{\mu\nu})^2 
+\frac{M^{-2\epsilon}}{2\xi\bar{g}^2}(\partial A)^2
\right) +\partial_\mu \bar c^a \left(D^{\mu}c\right)^a
\end{equation} 
where bearing in mind that $g_{*}^2$ $\sim$ $1/\Nf$ we have set $g^2$ $=$ 
$\bar g^2/\Nf$. Assuming that $\Nf$ is a large parameter, one can evaluate the 
integral with action~(\ref{Seff}) by the saddle point method that generates the
systematic expansion for the correlators. 

We can now examine which terms in~(\ref{Seff}) contribute to the leading IR 
asymptotics of the correlators. We start our analysis with the gluon propagator 
in the Landau gauge, $\xi$~$=$~$0$, which in leading order in $1/\Nf$ is 
\begin{equation}
G_{\mu\nu}(p) ~=~ \Nf^{-1}\biggr[a (p^2)^{d/2-1} + 
\frac{M^{-2\epsilon}}{2\bar{g}^2}p^2\biggl]^{-1} P_{\mu\nu}^{\perp} ~=~ 
\Nf^{-1}a\,(p^2)^{1-d/2}\left[ 1+\frac{1}{2a\bar{g}^2}
\left(\frac{p^2}{M^2}\right)^{\epsilon}\right]^{-1}P_{\mu\nu}^{\perp} ~. 
\end{equation}
Since $2\epsilon$ $=$ $4$ $-$ $d$ $>0$ the contribution to the propagator from 
the second term in the brackets of~(\ref{Seff}), $(F^a_{\mu\nu})^2$, is 
suppressed by $\left({p}/{M}\right)^{2\epsilon}$ in the IR region as compared 
with the contribution form the fermion loop which is the first term in the 
brackets. Thus the asymptotic form of the gluon propagator in the IR limit is 
fully determined by the fermion loop contribution
\begin{equation}
G_{\mu\nu}(p) ~\stackrel{IR}{\simeq}~ \frac{a}{\Nf}\, (p^2)^{1-d/2} 
P_{\mu\nu}^{\perp} ~.
\label{asform}
\end{equation}
Careful analysis shows that diagrams with triple and quartic gluon vertices do 
not contribute to the leading IR asymptotics of the correlators either. To see 
this we note at first that the term $(F^a_{\mu\nu})^2$ in the 
action~(\ref{Seff}) may be considered as an ultraviolet regulator with $M$ 
playing the role of a cutoff. Indeed, due to this term the gluon propagator 
decreases as $1/p^2$ as $p\to\infty$, which makes all diagrams convergent. We 
note that the appearence of divergences of the type $A_\mu^2$ is prohibited by 
gauge invariance. Next, if we let $G$ denote an arbitrary diagram, then 
rescaling the variables in the momentum integrals, $l_i\to M l^\prime_i$, one 
finds that up to some prefactor $M^{k\epsilon}$, the parameter $M$ enters the 
integrand together with the external momenta only. In other words 
$G(q_{ext},M)$ $=$ $M^{k\epsilon}G(q_{ext}/M)$. Therefore the limit 
$q_{ext}\to 0$ corresponds to the limit $M\to \infty$, which can be regarded as
the removal of the regularization. Further, since the most IR singular 
contributions we are interested in come from the integration over the region 
where $l^\prime_i\sim 0$, one can replace the gluon propagator by its 
asymptotic form~(\ref{asform}). Then it can be easily checked by simple 
dimensional analysis that the leading IR singularities come from the diagrams 
without gluon self-interaction vertices. On the formal level this follows from 
the fact that term~$(F^a_{\mu\nu})^2$ is accompanied by the factor 
$M^{-2\epsilon}$ and vanishes in the $M\to\infty$ limit. So one concludes that 
this term does not influence the critical properties of the theory and 
according to the general scheme \cite{Zinn} should be excluded from the action.
Therefore we obtain the theory defined by the action
\begin{equation}
\label{NATM}
S_{NATM} ~=~ \bar\psi\,(\!\not\!\partial-i\!\not\!\!A^{a}T^{a})\,\psi ~+~ 
\frac{\Nf}{2\xi}(\sqcap\!\!\!\!\!\sqcup^{-\epsilon/2}\partial A)^2 ~+~ 
\partial_\mu \bar c^a \partial^\mu c^a ~+~  
f^{abc}\partial^\mu \bar c^a A_\mu^b c^c ~,
\end{equation}
which in the Landau gauge has the same critical behaviour as QCD to all orders
in the $1/\Nf$ expansion. Of course, for gauge independent quantities it is 
true in any gauge. We have adapted the usual form of the gauge fixing condition
to ensure that the transverse and longitudinal parts of the gluon propagator 
have the same momentum dependence. It is interesting to note that in order to 
arrive at~(\ref{NATM}) one can start from the theory with manifest gauge 
invariance which is determined by the action $S$ $=$ $\bar\psi( \partialslash
- i\Aslash^{a}T^{a})\psi$ with ghost and gauge fixing terms in turn arising 
from the application of the Faddeev-Popov procedure to the functional integral. 

Power counting shows that the theory~(\ref{NATM}) is renormalizable within the 
$1/\Nf$ expansion and the renormalized action has the form 
\begin{equation}
\label{SR}
S_R ~=~ Z_1 \bar\psi\,\!\not\!\partial\psi ~-~
i Z_2\bar\psi\!\not\!\!A^{a}T^{a}\,\psi ~+~ 
\frac{\Nf}{2\xi}(\sqcap\!\!\!\!\!\sqcup^{-\epsilon/2}\partial A)^2 ~+~ 
Z_3\partial_\mu \bar c^a \partial^\mu c ~+~
Z_4f^{abc}\partial^\mu \bar c A_\mu^b c^c  
\end{equation}
where we assume, of course, that a gauge invariant regularization is used. Due 
to the Slavnov-Taylor identities the renormalization constants $Z_i$ are
related by 
\begin{equation}
\label{ST}
Z_2\,Z_1^{-1} ~=~ Z_4\, Z_3^{-1} ~.
\end{equation} 
This equation was used in the exponent formulation to determine the ghost 
anomalous dimension at $O(1/\Nf)$ in the Landau gauge in \cite{JG93}. As was 
proved above in the Landau gauge, the critical properties of QCD and this new 
theory which we shall refer to as the non-abelian Thirring model, (NATM), are 
identical. Therefore one can use the NATM to deduce the QCD RG functions. This 
equivalence at leading order in $1/\Nf$ was noted in \cite{H2} and used to
deduce various exponents at $O(1/\Nf)$ in 
\cite{JG91,JG92,JG94,JG93m,JGJB,JG93}. The extension of these calculations to 
$O(1/\Nf^2)$ requires special care. The main one is the necessity of using a 
gauge invariant regularization which was not crucial at $O(1/\Nf)$. The 
conventional dimensional regularization is not applicable here, since the gluon
propagator behaves as $(p^2)^{1-d/2}$ and the theory remains logarithmically 
divergent in any dimension $d$. To our knowledge most other invariant 
regularizations such as higher derivatives spoil the masslessness of the 
propagators, which makes higher order calculations virtually impossible. 
Usually in $1/N$ calculations the analytical regularization of \cite{VPH} is 
used. However, this breaks gauge invariance.

We now consider how to reconcile gauge invariance with the calculational 
advantage of using massless propagators. First, we break the gauge invariance 
of (\ref{NATM}) from the beginning by introducing a new coupling, $\lambda$, 
for the ghost-gluon vertex of (\ref{NATM}). For $\lambda$ $=$ $1$ we recover 
the original model but the theory remains renormalizable for arbitrary 
$\lambda$ as well. The only effect will be that the identity (\ref{ST}) will no 
longer hold. The bare coupling $\lambda_0$ is related to the renormalized one, 
$\lambda$, by
\begin{equation}
\label{lam}
\lambda_0 ~=~ Z_\lambda \lambda ~=~ Z_4\,Z_1\,Z_2^{-1}Z_3^{-1} \lambda ~,
\end{equation}
where the $Z_i$ will now also depend on $\lambda$. Let us suppose that we used 
an invariant method, such as regularization by higher derivatives \cite{HD}, to 
regularize this extended theory. Then it immediately follows from~(\ref{ST})
and (\ref{lam}) that the equality $\lambda$ $=$ $1$ for the renormalized 
coupling implies that $\lambda_0$ $=$ $1$ as well. Therefore we conclude that 
$\lambda$ $=$ $1$ is a fixed point, $\beta_\lambda(1)$ $=$ $0$. The existence 
of this fixed point is the key point and it does not depend on the 
regularization used. So using any other regularization can only change the 
position of the fixed point with in general $\lambda_{*}$ $=$ $1$ $+$ 
$O(1/\Nf)$. What is important, though, is that the anomalous dimensions 
calculated at the critical point, $\gamma(\lambda_{*})$, are scheme independent 
and, hence, coincide with the anomalous dimensions deduced in the original 
model~(\ref{NATM}). Therefore one can use the regularization which is most 
convenient from the computational point of view. Moreover, since we do not need
to consider diagrams with external ghost legs, then the only diagrams depending
on $\lambda$ are those with a ghost loop. As is evident from counting powers
of $1/\Nf$ these are themselves $O(1/\Nf^2)$. So at this order it is sufficient 
to set $\lambda$ $=$ $1$. 

\sect{Methods.}

As was shown above the RG functions of QCD at the critical points can be
deduced from the more simple NATM. The use of the $\Delta$-regularization 
\cite{VPH,VN82} allows us to retain massless propagators which means that the 
calculation of the higher order Feynman integrals can be achieved. However, the
price to be paid for this is the loss of the property of multiplicative 
renormalizability. Indeed, formally, the regularized action has the form
\begin{eqnarray}
\label{Sdelta}
S_\Delta&=& \bar\psi\!\not\!\partial\psi ~-~ i \bar\psi\!\not\!\!A^{a}T^{a}\psi
{}~+~ \frac{\Nf}{2\xi M^{2\Delta}}
\left( \sqcap\!\!\!\!\!\sqcup^{-(\epsilon-\Delta)/2}\partial^\mu A_\mu^a 
\right)^2 ~+~ \partial_\mu \bar c^a \partial^\mu c^a ~+~
\lambda f^{abc}\partial^\mu \bar c^a A_\mu^b c^c \nonumber \\
&&-~ \frac12A^a_\mu(x)M^{-2\Delta}K^{\mu\nu}_\Delta(x,y)A^a_\nu(y) ~+~ 
\frac12A^a_\mu(x)K^{\mu\nu}(x,y)A^a_\nu(y) ~.
\end{eqnarray}
We use $K^{\mu\nu}(x,y)$~$=$~$K^{\mu\nu}(x-y)$ for the inverse fermion loop in 
coordinate space with 
\begin{equation}
\left(K^{\mu\nu}(x)\right)^{-1} ~=~ -~ \frac{n\Gamma^2(\mu)}{4\pi^{2\mu}
|x^2|^{2\mu-1}} 
\left({g^{\mu\nu}-\frac{2x^\mu x^\nu}{x^2}}\right) ~, 
\end{equation}
where $n$~$=$~$\Nf \, T_F \, \mbox{Tr}_{\mbox{\footnotesize{spinor}}}I$. The 
regularized kernel $K^{\mu\nu}_\Delta(x,y)$ is defined by 
\begin{equation}
\label{regK} 
K^{\mu\nu}_\Delta(x) ~=~ C(\Delta)K^{\mu\nu}(x)|x|^{2\Delta}  
\end{equation}
and the constant $C(\Delta)$~$=$~$1$~$+$~$O(\Delta)$ will be specified below. 
The last term in~(\ref{Sdelta}) forms part of the interaction, while the 
penultimate one is assigned to the free part of the action. Since the last two 
terms in the action~(\ref{Sdelta}) are non-local, they are not renormalized and 
therefore the full renormalized action takes form
\begin{eqnarray}
\label{SRdelta}
S_\Delta^R &=& Z_1\bar\psi\,\!\not\!\partial\psi ~-~
i Z_2 \bar\psi\!\not\!\!A^{a}T^{a}\,\psi ~+~ 
\frac{\Nf}{2\xi M^{2\Delta}}
\left(\sqcap\!\!\!\!\!\sqcup^{-(\epsilon-\Delta)/2}\partial^\mu A_\mu^a 
\right)^2 ~+~ Z_3 \partial_\mu \bar c^a \partial^\mu c^a \nonumber \\ 
&& +~ Z_4 \lambda f^{abc}\partial^\mu \bar c^a A_\mu^b c^c ~-~ 
\frac12 A^a_\mu(x)M^{-2\Delta}K^{\mu\nu}_\Delta(x,y)A^a_\nu(y) ~+~ 
\frac12 A^a_\mu(x)K^{\mu\nu}(x,y)A^a_\nu(y) ~. \nonumber \\ 
\end{eqnarray} 
Obviously, $S_\Delta^R$ cannot be brought into the form~(\ref{Sdelta}) by the 
redefinition of the fields and constants. Consequently, the theory with 
action~(\ref{Sdelta}) is not multiplicatively renormalizable. A more detailed 
discussion of this topic can be found in the \cite{VN82,VS93}. The absence of 
multiplicative renormalizability means the following. We recall that in 
multiplicatively renormalizable theories the Green functions calculated within 
two different subtraction schemes are related to each other as follows
\[
G_n^I(x_1,\ldots,x_n,g) ~=~ A_nG_n^{II}(x_1,\ldots,x_n,Z_g\>g) ~,
\]
where the coefficient $A_n$ $=$ $Z_{\Phi}^n$ and $Z_{\Phi}$ is the field 
renormalization constant. In the case under consideration this equality holds 
as well. However, $A_n$ $\neq$ $Z_{\Phi}^n$ any longer with $A_n$ depending on 
$n$ in a nontrivial way. This of course does not contradict the statement about
the IR equivalence with QCD, since the IR asymptotics remains unchanged. 
Nevertheless, the absence of multiplicative renormalizability prevents us from 
using the standard method of RG analysis for the determination of critical 
exponents. The more widely used method for this is the method of 
self-consistency equations, which is based on the direct solution of the 
Schwinger-Dyson (SD) equations, \cite{VPH,VN82,JG91,JG92}. Another approach has 
been developed recently in \cite{DM,DMrus98} and will be briefly discussed 
below in the context of QCD.  

\subsection{Extended model.}
The convenience of the RG method from the computational point of view consists 
of the possibility of calculating critical exponents via the renormalizaton 
constants given in (\ref{RGf}). To retain this calculational advantage, 
following the lines of \cite{VN82,DM,DMrus98}, we restore the multiplicative 
renormalizability of the model in question by attaching two additional charges,
$u$ and $v$, to the last two terms of the action~(\ref{Sdelta}). Namely, we 
consider the model with the action
\begin{equation} 
\label{UVaction}
S_{uv} ~=~ S_{NATM} ~-~ \frac{u}{2}A^a_\mu(x)M^{-2\Delta}K^{\mu\nu}_\Delta(x,y)
A^a_\nu(y) ~+~ \frac{v}{2}A^a_\mu(x)K^{\mu\nu}(x,y)A^a_\nu(y) ~.
\end{equation}
\vspace{-3cm}
\begin{figure}[hb]
\centerline{\epsfxsize12.0cm\epsfbox[130 150  480 300]{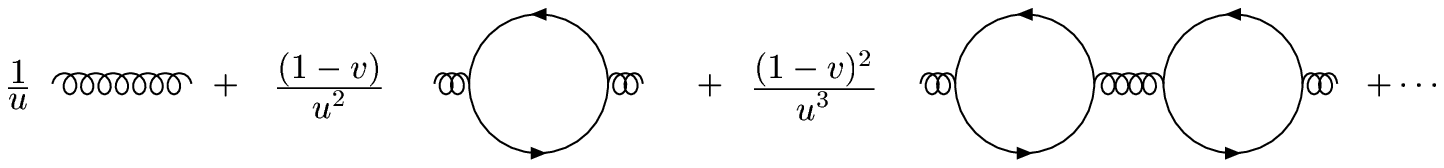}}
\caption[]{The effective gluon propagator for $u,v\neq 1$.}
\label{jsum}
\end{figure}
Obviously, the initial model~(\ref{Sdelta}) is recovered by the special choice 
of the parameters $u$~$=$~$v$~$=1$. As distinct from the initial model the 
propagator of the gluon field in the model~(\ref{UVaction}) with $u$ and $v$ 
couplings has a more complicated form. Indeed, at $v$~$\neq$~$1$ the last term 
in the action~(\ref{UVaction}) does not ensure the exact cancellation of the 
simple fermion loop insertion in the gluon line and one should sum up all such 
insertions~(see figure \ref{jsum}), that yield for the transverse part of the 
gluon propagator 
\begin{equation}
\label{line}
G^{\perp}(p;u,v) ~\equiv~ \frac{1}{u} G^{\perp}(p) \left(1 ~+~ \frac{(v-1)}{u}
t^{\Delta} ~+~ \frac{(v-1)^2}{u^{2}} t^{2\Delta} ~+~ \ldots \right) ~,
\end{equation}
with $t$~$\equiv$~$C(\Delta)(M^{2}/p^{2})$. This theory is obviously 
renormalized. In addition to the standard renormalization constants, $Z_1$, 
$\ldots$, $Z_4$, which depend now on the couplings $u$ and $v$, one should add
two new ones $Z_u$ and $Z_v$ to take account of the renormalization of the 
couplings $u$ and $v$,
\[
u_0 ~=~ u M^{-2\Delta} Z_u,\ \ v_0 ~=~ v Z_v,\ \ \  Z_u ~=~ Z_v ~=~ Z_A^{-2} ~.
\]
Being multiplicatively renormalizable this model can be analysed with standard 
RG methods. In particular the basic RG functions are 
\begin{eqnarray}
\label{explicitRG}
&&\gamma_{\Phi} ~=~ M\partial_{M}\ln{Z_{\Phi}} ~, \ \ \ 
\beta_{u} ~=~ M\partial_{M}u ~=~ 2u\>(\Delta -\gamma_{A}) ~, \ \ \ 
\beta_{v} ~=~M\partial_{M}v ~=~ -2v\gamma_{A} ~, \nonumber \\
&&\beta_\xi ~=~ M\partial_{M}\xi ~=~ -2\xi\>(\Delta -\gamma_{A}) ~, \ \ \ 
\beta_\lambda ~=~ M\partial_{M}\lambda ~, 
\end{eqnarray}
and
\begin{equation}
Z_\psi ~=~ Z_1^{1/2} ~, \ \ Z_c ~=~ Z_3^{1/2} ~, \ \ Z_A ~=~ Z_2\>Z_1^{-1} ~,
\ \ Z_\lambda ~=~ Z_4\>Z_3^{-1}Z_1\>Z_2^{-1} ~, \ \ Z_\xi ~=~ Z_A^{2} ~.
\end{equation}
Before discussing the question of the relation of the RG functions of the 
extended model to the critical exponents of QCD, we derive a compact expression
for the RG functions at the point $u$~$=$~$v$~$=1$. Henceforth we will adopt 
the minimal subtraction (MS) scheme in which the general renormalization 
constants have the form
\[
Z ~=~ 1 ~+~ \sum_{n=0}^{\infty} \frac1{\Delta^n}\>Z^{(n)}(u,v,\xi,\lambda) ~.
\]
Then taking into account the finiteness of the RG functions and using
(\ref{explicitRG}) one obtains 
\begin{equation}
\gamma ~=~ M\partial_{M}\ln{Z}|_{u=v=1} ~=~ 2(u\partial_u-\xi\partial_\xi) 
Z^{(1)}(u,v,\xi,\lambda)|_{u=v=1} ~.
\label{rgex}
\end{equation}
Since there are no derivatives with to respect $\lambda$ and $v$, we can set 
$\lambda$~$=$~$\lambda_*$~$=$~$1$~$+$~$O(1/\Nf)$ and $v$~$=$~$1$ from the very 
beginning. In this case only the first term in~(\ref{line}) survives and, 
therefore, the operation $(u\partial_u-\xi\partial_\xi)$ applied to a diagram 
gives simply the number of the gluon lines, $n_A$, in the latter
\begin{equation}
(u\partial_u-\xi\partial_\xi)\Gamma(u,v,\xi,\lambda)|_{u=v=1} ~=~ -~ 
n_A\Gamma(1,1,\xi,\lambda) ~.
\label{uva}
\end{equation} 
The formula~(\ref{rgex}) can be rewritten as follows
\begin{equation}
\gamma ~=~ -~ 2\sum_{\Gamma}n_A^{(\Gamma)} Z_\Gamma^{(1)}(1,1,\xi,\lambda) ~,
\label{nline}
\end{equation}
where the sum runs over the set of all diagrams. Thus to calculate the RG 
functions defined by (\ref{rgex}) one can put $u$~$=$~$v$~$=1$ from the 
very beginning but take into account the number of gluon lines in each 
individual graph. To represent (\ref{nline}) in a compact form we attach 
the factor $g$ to the gluon propagator 
\begin{equation}
G_{\mu\nu} ~\to~ g G_{\mu\nu} ~,
\label{mG}
\end{equation}      
so that each diagram with $k$ internal gluon lines acquires a factor $g^k$. 
Then for the anomalous dimensions of the basic fields, (\ref{rgex}) takes the 
form
\begin{equation}
\label{sf}
\gamma_\Phi ~=~ -~ \left. 2g\partial_g Z_\Phi^{(1)} 
\right|_{g=1} ~,
\end{equation} 
where the renormlization constants $Z_\Phi$ are calculated within the model
with $u$~$=$~$v$~$=1$ and the modified gluon propagator~(\ref{mG}).

Similarly, the matrix of anomalous dimensions of a system of composite 
operators, $\{O_i\}$, which mix under renormalization, is given by
\begin{equation}
\label{so}
\gamma_{ik} ~=~ \left. 2g\partial_g Z_{ik}^{(1)} 
\right|_{g=1} ~+~ \delta_{ik} n_{k,\Phi}\gamma_\Phi ~,
\end{equation}
where the mixing matrix, $Z_{ik}$, is defined in the standard way from the 
condition for the Green functions of the renormalized operators $O_{i}^R$, 
\begin{equation} 
O_{i}^{R} ~=~ Z_{ik}O_{k} ~,
\end{equation} 
to be finite. Again, $Z^{(1)}_{ik}$ is the coefficient of the simple pole in 
$\Delta$. One can easily note the obvious resemblance of the 
formul{\ae}~(\ref{sf}) and~(\ref{so}) with those used in the MS scheme in 
dimensional regularization. Therefore using this approach gives a simple and 
transparent way to compute the critical exponents of operators and, 
importantly, all the machinery of $\epsilon$ expansions can be adapted for the 
case in question as well.

\subsection{Critical exponents and RG functions.}

In the previous subsection we have developed the effective tool for the 
computation of the RG functions $\gamma_{RG}$~$=$~$\gamma_{RG}(u,v)|_{u=v=1}$ 
in the NATM. The non-trivial exercise is to show that these RG functions, 
coincide with the corresponding critical exponents $\gamma_{crit}$, which 
determine the scaling properties of the correlators. Generally speaking, this 
is not so and $\gamma_{RG}$~$\neq$~$\gamma_{crit}$. Indeed,  
the RG equation for 1-irreducible $n$-point Green functions with an operator 
insertion in the extended $uv$ model reads
\begin{equation}
\Bigl( \left [ M\partial_{M}+\beta_{U}\partial_U+\beta_\xi\partial_\xi
+\beta_\lambda\partial_\lambda-n_{\Phi}\gamma_{\Phi} \right] \delta^{ik} 
+\gamma^{ik}_{RG} \Bigr ) \Gamma_k(p_{1},\ldots,p_{n+1};u,v,\xi,\lambda) ~=~ 
0 ~.
\label{rgu0}
\end{equation} 
Bearing in mind the critical equivalence between QCD and the NATM in the Landau
gauge, we set $\xi$~$=$~$0$ and $\lambda$~$=$~$\lambda_*$ and do not display 
their explicit dependence in the Green functions. Since $\beta_\xi(0)$ $=$ 
$\beta_\lambda(\lambda_{*})$ $=0$ from (\ref{explicitRG}), the above equation 
takes the form
\begin{equation}
\Bigl( \left [ M\partial_{M}+\beta_{U}\partial_U-n_{\Phi}\gamma_{\Phi}
\right]\delta^{ik}+\gamma^{ik}_{RG} \Bigr )
\Gamma_k(p_{1},\ldots,p_{n+1};u,v,\xi,\lambda) ~=~ 0 ~.
\label{rgu}
\end{equation} 
Evidently, due to the presence of the term $\beta_U\partial_U$ in (\ref{rgu}) 
the latter does not describe the scaling properties of the Green function 
$\Gamma_k(p_{1},\ldots,p_{n+1};u,v)$ under scale transformation even at the
point $u$~$=$~$v$~$=1$. To make this more transparent, taking into account 
(\ref{explicitRG}), we rewrite (\ref{rgu}) as follows 
\begin{equation}
\label{anterm}
\Bigl( \left [ M\partial_{M}-n_{\Phi}\gamma_{\Phi} \right]\delta^{ik} 
+ \gamma^{ik}_{RG} \Bigr) \Gamma_k(\{p_{l}\}) ~=~ 
2\delta^{ik}\gamma_A(\partial_u+\partial_v)
\Gamma_k(\{p_{l}\};u,v)\Bigl|_{u=v=1} ~.
\end{equation} 
It is evident that the matrix $\gamma^{ik}_{RG}$ can be considered as the 
matrix of the anomalous dimensions only if the right side of (\ref{anterm})
is equal to zero. However, since in the Landau gauge $\gamma_A\neq0$ the right 
side does not vanish identically. This means that knowledge of the matrix 
$\gamma^{ik}_{RG}$ alone is not sufficient for the calculation of the critical 
exponents.

The problem of the relation of the $\gamma_{RG}$ and $\gamma_{crit}$ has been 
analysed in \cite{DM,DMrus98} where it had been shown for the example of the 
$O(N)$ nonlinear sigma model that the difference 
$\gamma_{RG}$~$-$~$\gamma_{crit}$ is $O(1/N^3)$. Thus, up to $O(1/N^2)$ the 
formul{\ae}~(\ref{sf}) and (\ref{so}) give the true answer for the critical 
exponents. To prove this statement one has to show that the right side of 
(\ref{anterm}) is $O(1/N^3)$. Since the anomalous dimension of the gluon field 
is $O(1/\Nf)$, it is sufficient to prove that the quantity 
$(\partial_u+\partial_v) \Gamma_k(\{p_{l}\};u,v)\Bigl|_{u=v=1}$ vanishes at 
$O(1/\Nf)$ for any Green function. The proof given in \cite{DM} is reduced
to the problem of checking this property and without any change can be simply 
adapted to the present model. More details can be found in \cite{DM,DMrus98}. 
Thus we have justified the use of formul{\ae}~(\ref{sf}) and (\ref{so}) for the 
computation of critical exponents in the NATM or QCD in the Landau gauge up to 
$O(1/\Nf^2)$.

It is worth considering what happens in other gauges. Of course, as has been
stressed earlier, in this case it is only sensible to consider the anomalous
dimensions of gauge invariant operators. In general, the anomalous dimensions 
of gauge invariant operators, such as $\bar\psi\psi$ or $(F^a_{\mu\nu})^2$, 
calculated with (\ref{sf}) and (\ref{so}) in the extended model, will be 
gauge dependent. Indeed, we broke the gauge invariance of the model explicitly 
by introducing the coupling $\lambda$ for the ghost-gluon vertex and implicitly
by the regularization we have introduced. So it is not surprising that the 
correlators of gauge invariant objects in this model appear to be gauge 
dependent. However, it follows immediately from the critical equivalence of QCD
and the NATM that at the critical point $\lambda$~$=$~$\lambda_*$ all 
dependence on $\xi$ in the correlators of gauge invariant operators factorizes
\begin{equation}
\label{GG}
\Gamma(x_i,\xi) ~=~ A_\Gamma(\xi) \tilde \Gamma(x_i) ~,
\end{equation}
where $\tilde \Gamma(x_i)$ is already independent of $\xi$. Then the RG 
equation for the correlator of two gauge invariant operators $O_1(x)$ and 
$O_2(y)$ can be cast in the form
\begin{equation}
\label{gx}
\Bigl( M\partial_{M}+\gamma_{O_1}+\gamma_{O_2}+\gamma^* \Bigr) 
\tilde\Gamma(x,y) ~=~ 0 ~,
\end{equation}  
where 
$\gamma^*$~$=$~$\beta_\xi\partial_\xi \ln{A(\xi)}$. Bearing in mind that the 
explicit breaking of gauge invariance occurs at $O(1/\Nf)$ only, 
($\lambda_*$~$=$~$1$~$+$~$O(1/\Nf)$), one can deduce that $\gamma^*$ is 
$O(1/\Nf^3)$. This immediately leads us to the conclusion that up to 
$O(1/\Nf^2)$ the RG functions of the gauge invariant operator given by the 
formul\ae~(\ref{sf}) and (\ref{so}) do not depend on the gauge fixing 
parameter.

Finally, since $\beta_u$ and $\beta_v$ as well as $\beta_\xi$ are proportional
to $\gamma_A$ from (\ref{explicitRG}), one can nullify all these beta functions
simultaneously choosing the value of the gauge fixing parameter from the 
condition
\begin{equation}
\label{xi_A} 
\gamma_A(\xi_A) ~=~ 0 ~, \ \ \ \ \ \ \ 
\xi_A ~=~ -~ \frac{(2\mu-1)}{(\mu-1)}~+~O\left(\frac{1}{\Nf}\right) ~.
\end{equation} 
In this gauge the RG equation in the extended $uv$ model takes the desired form
\begin{equation}
\Bigl( \left[ M\partial_{M}-n_{\Phi}\gamma_{\Phi} \right]\delta_{ik} 
+\gamma^{ik}_{RG} \Bigr) \Gamma_k(\{p_{l}\}) ~=~ 0 ~,
\end{equation} 
from which it follows that the RG functions $\gamma_{RG}$ are the genuine 
anomalous dimensions $\gamma_{crit}$. 

We conclude this section by summarizing the main results:
\begin{itemize}
\item In the Landau gauge the critical exponents of QCD at the critical point 
can be calculated with standard RG methods via the renormalization constants in
the extended NATM with the help of the formul{\ae}~(\ref{sf}) and (\ref{so}) up
to $O(1/\Nf^2)$.
\item Up to $O(1/\Nf^2)$ the correspondence between the critical exponents of 
gauge invariant operators in QCD and NATM holds in any gauge.
\item For the special value of the gauge fixing parameter $\xi$~$=$~$\xi_A$ the
formul\ae~(\ref{sf}) and (\ref{so}) yield the anomalous dimensions of the gauge 
invariant operators in {\it all orders} of $1/\Nf$ expansion. 
\end{itemize}
In the next sections we will demonstrate the effectiveness of the above 
approach by calculating the anomalous dimensions of the quark field and the 
mass operator $\bar\psi\psi$ at second order in the $1/\Nf$ expansion. The 
explicit values of the renormalization constants $Z_i$ and the basic RG 
functions at $O(1/\Nf)$ are given in appendix B. 

\sect{Calculation of graphs.}

In this secton we discuss the technical details of the calculation of the 
diagrams relevant for the determination of the anomalous quark dimension. The 
technique for the evaluation of massless Feynman diagrams is widely documented.
(See, for example, \cite{VPH82,Kazakov}). Nevertheless, for completeness, we 
provide a list of basic formul\ae, such as the rules for the integration of 
chains of propagators for scalar, spinor and vector fields and the uniqueness 
relation for the different type of vertices in appendix A. In what follows we 
shall concentrate mainly on the calculational problems specific to the case 
under consideration.

\begin{figure}[t]
\centerline{\epsfxsize12.0cm\epsfbox[100 250  500 400]{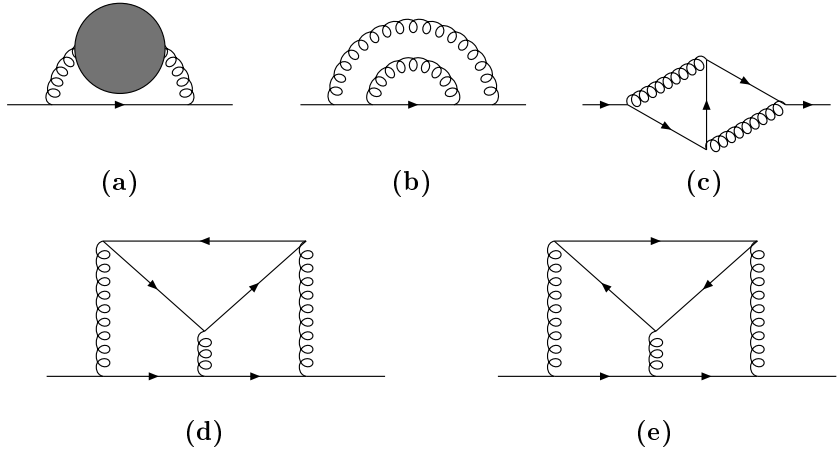}}
\vspace*{1cm}
\caption[]{Diagrams contributing to the computation of $\eta_2$. The first
graph represents the gluon self energy diagrams of figure 
\protect{\ref{fig2}.}} 
\label{fig1}
\end{figure}

First, we consider the diagrams with external quark legs which are illustrated 
in figure \ref{fig1}. It transpires that these are not as complicated to 
evaluate as may appear at first sight. Indeed diagrams (a) and (b) are trivial
in that they are equivalent to simple chain graphs in the language of 
\cite{VPH}. Although (c) has been considered in the case of QED we re-evaluate
it here in the context of the method used to determine diagrams (d) and (e)
where the latter differ from each other by a colour factor. 

\begin{figure}[ht]
\centerline{\epsfxsize12.0cm\epsfbox[200 250  500 320]{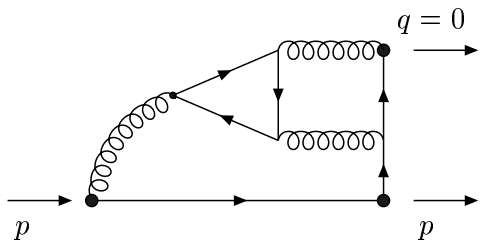}}
\caption[]{New external momenta routing for the diagram 
Fig.~\protect{\ref{fig2}},~(d).}
\label{expl}
\end{figure}

First, for (c) and (d) it is convenient to choose the flow of the external 
momentum $p$ in both diagrams to be along the fermion line connecting the 
external vertices. The diagrams are linearly divergent. Differentiating them 
with respect to the external momentum $p_\mu$ one obtains a set of 
logarithmically divergent diagrams, which differ from the initial ones by the 
insertion of a new vertex $(-\gamma_\mu)$ in each fermion line. This is due to 
the result 
\begin{equation} 
\frac{\partial}{\partial p^\mu} \, \frac{\pslash+\qslash}{(p+q)^2} ~=~ -~ 
\frac{\pslash+\qslash}{(p+q)^2}\>\gamma_\mu\> \frac{\pslash+\qslash}{(p+q)^2}~.
\end{equation}
We recall that after subtraction of the divergent subgraphs, in other words 
after the application of the ${\cal R}^\prime$ operation, the residues of the 
poles in $\Delta$ in the diagrams do not depend on the external momenta. 
Therefore, we are free to choose an arbitrary route for the flow of the 
external momenta in the each of the resulting diagrams to simplify the 
subsequent calculations. Our choice is the following. For diagram (c) with the 
insertion of the vertex $(-\gamma_\mu)$ in the vertical fermion line we direct 
the external momenta flow from the upper to the bottom vertices. When the 
insertion is in the right or left fermion line we choose the route of the 
external momenta flow from this new vertex to be along the fermion line nearest
to the incoming (outgoing) vertex. After this rearrangement all diagrams are 
easily integrated since they are reduced to simple chains. For diagram~(d) we 
choose the inserted vertex adjacent to the external one as the new external 
vertex, and direct the momentum along the fermion line connecting them. Then, 
the resulting diagram is also reduced to chains but with the insertion of the 
two loop master diagram in one of the lines, see figure~\ref{expl}. The 
evaluation of this two loop diagram is straightforward.  

\begin{figure}[hb]
\centerline{\epsfxsize10.0cm\epsfbox[100 250  500 450]{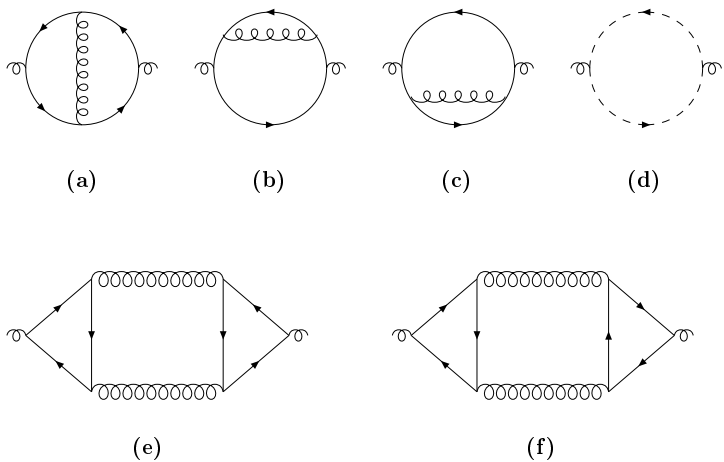}}
\vspace{1cm}
\caption[]{The diagrams contributing to the gluon self-energy at $O(1/\Nf^2)$.}
\label{fig2}
\end{figure}

We now turn to the discussion on the calculation of the diagrams contributing
to the gluon self-energy. These are shown in figure \ref{fig2} where (b), (c) 
and the graph involving ghosts, (d), are again trivial to determine and do not
deserve further comment. On the contrary graphs (a), (e) and (f) are tougher
to evaluate and we give details of their determination. As (e) and (f) are  
related by the same up to the colour factor we will focus on the former. Again 
(a) was evaluated in the QED calculation but we reconsider it here due to the 
fact that new techniques were required to evaluate (e) which can more easily be
appreciated in a two loop topology. 
We recall first the standard method for calculating the diagrams with these 
topologies in the example of the nonlinear $\sigma$ model, \cite{VPH}. It will
allow us to avoid the unnecessary complications related to the nontrivial
$\gamma$-matrix structure of these diagrams in the case of QCD. So, we first
consider diagram~(a) of figure \ref{fig2}, where now all the lines are assumed 
to be scalar ones, with respective indices $(2-\Delta)$ and $(\mu-1)$ for the 
wavy and arrowed lines. The diagram is superficially convergent but has two 
divergent subgraphs. If one could set $\Delta$~$=$~$0$ then the diagram could 
be easily integrated due to the uniqueness of the $3$-point vertex. To 
determine this diagram it was suggested in \cite{VPH} to subtract and add 
diagrams which have the same singularities as the initial one, but which can be
explicitly computed.
\centerline{\epsfxsize12.0cm\epsfbox[100 400  500 250]{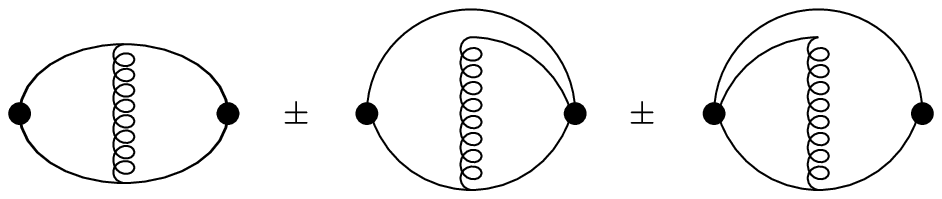}}
\vskip 1.5cm
\noindent
Since the singularities cancel in the difference one can take the limit 
$\Delta$~$\to$~$0$. Then all diagrams can be calculated due to uniqueness of 
the $3$-point vertex. The extension of this approach to the case of QCD is 
possible but leads to the certain problems. The first one is the increase in 
the number of diagrams to be evaluated. In other words five instead of the one 
for the nonlinear $\sigma$-model. The second and more difficult one is that for 
$\Delta$~$=$~$0$ the quark-gluon vertex is not unique. Using the terminology 
of \cite{Kazakov} it is one step from uniqueness, which suggests the 
application of the integration by parts method. This also increases the number 
of integrals to be considered. The calculation of diagrams with propagators 
having nontrivial tensor structure is more involved compared to scalar ones. 
Therefore it is desirable to try and keep the number of diagrams during a 
calculation to a minimum. The approach which is the most economical in this 
sense is likely to be the one advocated in \cite{Kazakov}. Again, we explain 
the idea in the simple example of the $\sigma$ model. We shift the indices of 
the upper lines in diagram~(a) by an amount $\epsilon$ and the lower ones by 
$-$~$\epsilon$. In other words we set $\mu-1$~$\to$~$\mu-1+\epsilon$ in the top
lines. It is important to note that here $\epsilon$ is a temporary 
regularization and ought not to be confused with the parameter which is 
conventionally used as the dimensional regularization in standard perturbative 
calculations. The analytic expression for such a diagram can now be written as
\begin{equation}
\frac{1}{\Delta}F(\Delta,\epsilon),
\end{equation}
where $F(\Delta,\epsilon)$ is a regular function in the vicinity of the point 
$\Delta$~$=$~$\epsilon$~$=$~$0$. What is important is that we have shifted the 
indices of the lines in such a way that the pole structure of the diagram has 
remained unchanged. Next, we need to know this diagram up to constant term in 
$\Delta$ at $\epsilon$~$=$~$0$. Due to the obvious symmetry 
$\epsilon$~$\to$~$-\epsilon$ the function $F$ only depends on $\epsilon^2$, and
its expansion near $\Delta$~$=$~$\epsilon$~$=$~$0$ is
\begin{equation}
F(\Delta,\epsilon) ~=~ F_0+\Delta F_1 ~+~ \Delta^2 F_3 ~+~ \epsilon^2 F_4 ~+~ 
O(\Delta^3,\Delta\epsilon^2,\epsilon^4) ~.
\label{Expans}
\end{equation}
Since the first two terms of the expansion we are interested in are independent
of $\epsilon$, one is allowed to evaluate the diagram at any value of 
$\epsilon$. It is convenient to set $\epsilon$~$=$~$\Delta/2$ which results in
the uniqueness of the upper vertex and the diagram is immediately integrable.
It is worth noting that one has only to calculate a single diagram instead of
five. Applying this procedure twice to the diagram with topology~(e) one 
reproduces the known answer \cite{VPH} with a minimum amount of calculation.
For further discussion it will be important that one is able to determine the 
subsequent, $\Delta^2$ term in the expansion~(\ref{Expans}) as well. Indeed, 
bearing in mind that the pole term $F(0,\epsilon)/\Delta$ is fully determined 
by the counterterm diagrams or, equivalently, is given by the application of 
the ${\cal R}^\prime$ operation to the graph in question, one can easily deduce 
the coefficient $F_4$. This in its turn allows us to fix the value of the 
coefficient $F_3$. 

It now seems reasonable to apply this idea to the QCD diagrams. However, one
obstacle remains in that the quark-gluon vertex is not unique for the choice
of exponents we are restricted to. Though when the propagator of the vector 
fields has the conformal form
\begin{equation}
G_{\mu\nu}^{conf}(x) ~=~ \frac{1}{(x^2)^\beta} \left( \eta_{\mu\nu} 
-\frac{2x_\mu x_\nu}{x^2} \right) ~,
\label{confprop}
\end{equation}
then the $3$-point fermion vector vertex is unique if the sum of the vertex 
indices, $2\alpha$~$+$~$\beta$, is equal to $d+1$, where $\alpha$ is the index
of the fermion line. For this unique vertex the relation given in appendix A 
holds. However the propagator of the gluon field does not have a conformal 
form. Indeed, at first order in the $1/\Nf$ expansion in the Landau gauge it 
reads 
\begin{equation}
\label{GLandau}
G_{\mu\nu}(x) ~=~ \frac{\tilde G}{(x^2)^{1-\Delta}} \left[ \eta_{\mu\nu}+
\frac{2(1-\Del)}{(2\mu-3+2\Del)} \frac{x_{\mu}x_{\nu}}{x^2} \right] ~,
\end{equation}
where $\tilde G$ is some constant. Of course, bearing in mind that the 
calculation of the diagrams with a longitudinal gluon propagator is rather 
trivial one may represent~(\ref{GLandau}) in the following form
\begin{equation}
\label{decomp}
G_{\mu\nu}(x) ~=~ A(\mu,\Delta)G_{\mu\nu}^{conf}(x) ~+~ 
B(\mu,\Delta)G_{\mu\nu}^{\parallel}(x) ~,
\end{equation}
where $A$ and $B$ are some constants and $G_{\mu\nu}^{\parallel}(x)$ is purely
longitudinal in momentum space. In such a decomposition it is easy to check 
that the constants $A$ and $B$ are singular as $\Delta$~$\to$~$0$, 
$A$~$\sim$~$B$~$\sim$~$1/\Delta$. The reason for this is that when 
$\Delta$~$=$~$0$ the conformal propagator~(\ref{confprop}) coincides with the 
longitudinal one. If these constants were finite then one could calculate the 
diagram for the conformal and longitudinal part of the propagator separately. 
The diagrams with longitudinal propagator would be trivial to integrate, whilst
for those with a conformal propagator could be evaluated with the methods we 
discussed earlier. Now, the singularity in the coefficients of $A$ and $B$ 
lead to additional difficulties. Namely, due to the additional factor 
$1/\Delta$ in the coefficients $A$ and $B$ one should evaluate this diagram 
with higher accuracy in $\Delta$. However it is a reasonable price to pay for 
the considerable simplification which arises from the uniqueness of the triple 
vertex. 

In the following we shall mainly discuss diagram~(e) of figure \ref{fig2} since 
its evaluation was the most difficult. We will focus our discussion on the more
important points as the intermediate steps are not difficult to reproduce. 
However, to give an impression about the effectiveness of the suggested 
appproach it is instructive to discuss the calculation of the QED type,  
diagram figure \ref{fig2}(a), first. Using the decomposition~(\ref{decomp}) one 
reduces the problem to the calculation of the diagram with the conformal 
propagator~(\ref{confprop}) with index $\gamma$ $=$ $1$ $-$ $\Delta$, where the 
calculation of the diagram with longitudinal propagator is trivial. Due to the 
singularity of the coefficient $A(\Delta)$ this diagram should be evaluated up 
to terms linear in $\Delta$. Following the scheme discussed above in the case 
of the scalar diagram we introduce the additional regularizing parameter $\pm 
\epsilon$ in the upper and lower lines. Then the analytic expression for this 
diagram is given by the same formula~(\ref{Expans}), where all the necessary 
Lorentz indices are implied. For $\epsilon$ $=$ $\Delta/2$ the diagram can be
integrated due to the uniqueness of the upper vertex, while the coefficient 
$F_4$, as was explained earlier, can be extracted from the one loop conterterm 
diagrams. Thus, we have reduced the determination of the photon self energy 
diagram to the calculation of four chains. 

We now consider the calculation of the three loop gluon self energy diagram.
\begin{figure}[ht]
\centerline{\epsfxsize12.0cm\epsfbox[140 240  500 360]{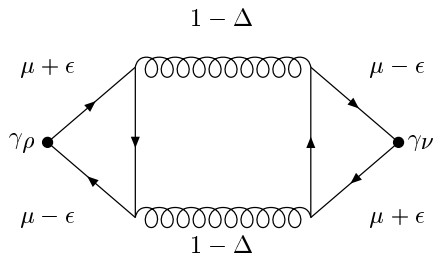}}
\caption[]{Three loop gluon self energy diagram with $\epsilon$ and $\Delta$ 
regularizations.}
\label{shifted}
\end{figure}
First, since the longitudinal part of the gluon polarization tensor 
$\Pi_{\rho\nu}(p)$ can be easily calculated, it is sufficient to calculate the 
above diagram with contracted indices, $\Pi_\nu^{~\nu}$, which simplifies the 
algebra significantly. Moreover, the contributions of all the diagrams in 
figure \ref{fig2} to the longitudinal part of the gluon polarization cancel 
identically. Further, since the calculation of the diagram with the 
longitudinal propagator is trivial, we shall discuss only those with the 
conformal ones. For this diagram we again introduce the additional 
regularization $\epsilon$ in the fermion lines as shown in figure 
\ref{shifted}. The diagram, being superficially convergent with two divergent 
two loop subgraphs, the analytic expression for it is given as follows
\begin{equation}
\Pi(\Delta,\epsilon) ~=~ \frac1\Delta \left(\Pi_0+\Delta \Pi_1+\Delta^2 
\Pi_2+\Delta^3 \Pi_3+ \epsilon^2 \Pi_4+\epsilon^2\Delta \Pi_5 
+ O(\Delta^4,\Delta^2\epsilon^2,\epsilon^4) \right) ~.
\end{equation}
We recall that our purpose is to determine the function $\Pi(\Delta,0)$ up to
the $\Delta^3$ terms. In other words we are interested in the coefficients 
$\Pi_0$, $\Pi_1$, $\Pi_2$ and $\Pi_3$. 

The strategy of the calculation is the following. For $\epsilon$ $=$ $\Delta$ 
one can exploit the uniqueness of the lower right and upper left vertices to 
evaluate the diagram, which allows us to find the coefficients $\Pi_0$ and 
$\Pi_1$ and the combination of coefficients $\Pi_2+\Pi_4$ and $\Pi_3+\Pi_5$. To 
extract $\Pi_2$ and $\Pi_3$ we evaluate the coefficients $\Pi_4$ and $\Pi_5$. 
The former is determined by the counterterm diagrams and its calculation is 
straightforward, while the evaluation of the latter is more nontrivial and will
be discussed below. At first, however, we consider in more detail the 
computation of the diagram for a special value of the parameter $\epsilon$. We
take $\epsilon$ $=$ $\Delta$, which restores the uniqueness of the two 
vertices.

\begin{figure}[t]
\centerline{\epsfxsize12.0cm\epsfbox[140 350  450 500]{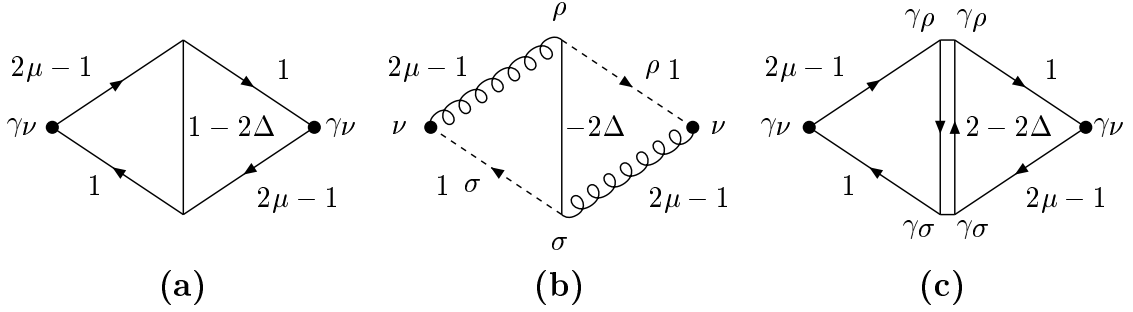}}
\vspace{-0.6cm} 
\caption[]{The three resulting diagrams for $G^{conf}\otimes G^{conf}$ case.}
\label{cc}
\end{figure}

Using the uniqueness relation one can express the initial diagram as the sum of
two loop diagrams, which after some algebra can be reduced to the ones shown in 
figure \ref{cc} and a chain integral. Here the wavy line with labels $\nu$ and 
$\rho$ denotes the conformal propagator $G^{conf}_{\nu\rho}$. The dashed line 
with label $\rho$ and index $\alpha$ is used for the propagator 
$x_\rho/(x^2)^\alpha$. In diagram~(c) the vertical double line is used to 
represent $\xslash\otimes\xslash$ where each $\xslash$ enters from a different 
fermion cycle.
We discuss briefly the calculation of each of the diagrams shown in figure 
\ref{cc}.
\begin{itemize}
\item{Diagram~(a).\\
This diagram enters the expansion with a coefficient proportional to
$\gamma-1$~$=$~$-\Delta$ so we need to calculate it up to $O(\Delta)$. To this 
end we shift the indices of upper fermion line by $\epsilon$, and the lower 
ones by $-\epsilon$. The analytic structure of this modified diagram is the
following
\begin{equation}
\Pi_a(\Delta,\epsilon) ~=~ \frac{R(\epsilon)}{\Delta} ~+~ 
R^\prime(\Delta,\epsilon) ~=~ \frac1\Delta(R_0+\epsilon^2 R_1 + 
O(\epsilon^4)) ~+~ R^\prime_0 ~+~ R^\prime_1\Delta ~+~ 
O(\Delta^2,\epsilon^2) ~,
\end{equation}
where we took into account the fact that due to the 
$\epsilon$~$\to$~$-\epsilon$ symmetry the diagram depends on $\epsilon^2$ only.
Further, the diagram can be integrated at $\epsilon$~$=$~$\Delta$ due to 
uniqueness of the Gross-Neveu type at the upper vertex. Again, the coefficient 
$R_1$ is determined by the counterterm diagrams and can be easily evaluated. 
This is sufficient to determine $\Pi_a(\Delta,0)$ with the required accuracy.   
}
\item{Diagram~(b).\\
At first we note that this diagram is finite as $\Delta$~$\to$~$0$ and we need 
it up to $O(\Delta^2)$. Again, we shift the indices of the wavy lines by 
$+\epsilon$ and $-\epsilon$. Then the expansion of this new diagram in $\Delta$
and $\epsilon$ reads
\begin{equation}
\Pi_b(\Delta,\epsilon) ~=~ \Delta\, R_0 ~+~ \Delta^2 R_1 ~+~ \epsilon^2 
R_2 ~+~ \ldots\ ~.
\end{equation}
When $\Delta$~$=$~$0$ the diagram can be calculated exactly which allows us to 
find the coefficient $R_2$. Next, it can be checked that the diagram can be
integrated for $\epsilon$~$=$~$2\Delta$ as well. Indeed, representing 
$(\eta_{\nu\rho}-2x_\nu x_\rho/x^2)(z-x)^\rho$ as
\[
z_\nu ~-~ x_\nu \frac{z^2}{x^2} ~+~ x_\nu \frac{(z-x)^2}{x^2} ~,
\] 
where we choose the coordinates of the left and right vertices to be $0$ and 
$z$ respectively, and the upper vertex to be $x$, one obtains three diagrams. 
The first two of them can be calculated due to the uniqueness of the upper 
vertex, whilst the last is a simple chain. We note that one needs to introduce 
an additional regularization $\delta$, by for example shifting the index of the
lower dashed line of the original diagram by $\delta$, to make each of the 
resulting diagrams finite. Of course, the sum of the diagrams is finite in the 
$\delta$~$\to$~$0$ limit. Again it is sufficient to determine $\Pi_b(\Delta,0)$
with the required accuracy, 
$\Pi_b(\Delta,0)$~$=$~$\Pi_b(\Delta,2\Delta)$~$-$~$\Pi_b(0,2\Delta)$.
}
\item{Diagram~(c).\\
The last diagram enters the expansion with the coefficient 
$(\gamma-1)^2$~$=$~$\Delta^2$. So one needs to calculate it with $O(1)$ 
accuracy. The calculation runs along the same lines as above. We shift the
indices of the upper lines by $+\Delta$, and the lower ones by $-\Delta$, which
evidently does not influence the $\Delta^0$ terms. To simplify the 
$\gamma$-matrix structure of the diagrams it is convenient to represent the
tensor product of the $\gamma$-matrix, $\gamma_\rho\otimes\gamma^\rho$ where
each $\gamma$-matrix enters from different traces, as 
$\gamma^\rho{\hat G}_{\rho \lambda} \otimes 
{\hat G}^{\lambda\rho^\prime}\gamma_{\rho^\prime}$, where $x$ is the coordinate
of the upper vertex and ${\hat G}_{\rho \lambda}$ is the numerator of the 
conformal propagator~(\ref{confprop}). We repeat this exercise for 
$\gamma_\sigma\otimes\gamma^\sigma$ and $\gamma_\nu\otimes\gamma^\nu$. Then, 
using the following identities
\[
\xslash \gamma^\rho {\hat G}_{\rho \lambda}(x) ~=~ - \, \gamma_\lambda 
\xslash ~, \ \ \ {\hat G}_{\lambda\sigma}(x)\gamma^\sigma ~=~ - \, 
\xslash\gamma_\lambda\xslash/x^2
\] 
both traces can be cast into the form suitable for the inversion 
transformation. After the inversion the upper vertex decouples from the base 
(left vertex) and the diagram can be immediately integrated.
}
\end{itemize}

\begin{figure}[t]
\centerline{\epsfxsize12.0cm\epsfbox[140 240 470 360]{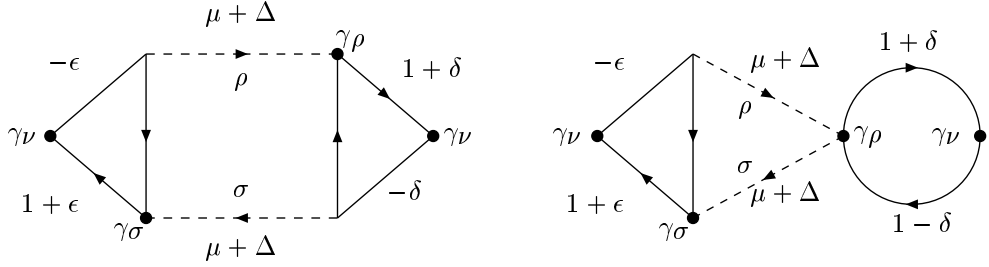}}
\caption[]{Two diagrams contributing to the evaluation of $\Pi_5$. The indices 
of lines are given in the momentum space version.}
\label{long}
\end{figure}

The evaluation of the coefficient $\Pi_5$ is grounded on the following
observation. Since in the momentum space representation the conformal 
propagator takes the form
\begin{equation}
G^{conf}(p) ~=~ \frac{\tilde A}{(p^2)^{\mu-1+\Delta}}
\left (P^{\parallel}(p)-\frac{\Delta}{2(\mu-1+\Delta)}P^{\perp}(p)\right )
\label{confp}
\end{equation}
the contributions to the $\epsilon^2\Delta$ term come only from the momentum
integral involving two longitudinal propagators $P^{\parallel}\otimes 
P^{\parallel}$, or longitudinal and transverse ones, 
$P^{\parallel}\otimes P^{\perp}$. Moreover, because the transverse propagator 
enters the expansion~(\ref{confp}) with an additional factor $\Delta$, the 
contribution from the latter can be extracted again from the consideration of 
the counterterm diagrams alone, and does not require much work. Thus the 
nontrivial part of the calculation is the determination of the diagram in 
question with the longitudinal propagators. We shall use the momentum 
representation and at the first step contract one momentum from the numerator 
of each longitudinal gluon propagators into the fermion traces. After this the 
initial diagram reduces to four integrals. Two of these are identical whilst 
another is a simple chain integral. Therefore, we are left with two nontrivial 
diagrams which are shown in figure \ref{long}. For later convenience we replace 
the regulator $\epsilon$ in the right fermionic triangles by $\delta$.  
We consider the left hand diagram first and denote it by 
$G(\Delta,\epsilon,\delta)$. We look for the $\epsilon^2$ term in the expansion
of function $G(\Delta,\epsilon,\epsilon)$. It is easy to see that for 
$\epsilon$~$\to$~$0$ with $\Delta$ and $\delta$ fixed that $G$~$\to$~$0$ as 
well. Indeed in the $\epsilon$~$\to$~$0$ limit the left hand triangle turns 
into a fermion loop, which is transverse, so that when it is contracted with
the incoming momenta it vanishes. Thus one concludes that 
$G(\Delta,\epsilon,\delta)$~$=$~$\epsilon 
\delta{\tilde G}(\Delta,\epsilon,\delta)$ and for our purpose it is therefore
sufficient to calculate $\epsilon\delta{\tilde G}(0,0,0)$, where we take into 
account the finiteness of the diagram. This implies the following method for 
its evaluation. We ascribe the label $1$ to the regulators $\epsilon$ and 
$\delta$ on the upper fermion lines and replace $\epsilon$~$\to$~$\epsilon_1$  
and $\delta$~$\to$~$\delta_1$. Similarly for the lower lines we replace 
$\epsilon$ by $\epsilon_2$ and $\delta$ by $\delta_2$ there. Hence, the diagram
now depends on four variables $G(\epsilon_1,\epsilon_2,\delta_1,\delta_2)$. 
Then $\epsilon^2{\tilde G}(0,0,0)$~$=$~$G(\epsilon,0,\epsilon,0)$ 
$+$~$G(\epsilon,0,0,\epsilon)+G(0,\epsilon,0,\epsilon,0)$ 
$+$~$G(0,\epsilon,0,\epsilon)$ and each of these four diagrams can be reduced 
to chain integrals which completes the evaluation of this graph.  

The calculation of the right hand diagram of figure \ref{long} is more 
involved. In momentum space it corresponds to
\begin{equation}
\Pi(\Delta,\epsilon) ~=~ \frac{\Pi^{\nu\rho}(\Delta,\epsilon) 
P^\perp_{\nu\rho}(p)}{(p^2)^{1-\mu+2\Delta}} ~,
\label{rightd}
\end{equation}
where the transverse projector arises from the fermion loop and the nontrivial
piece, $\Pi_{\nu\rho}$, comes from the two loop master diagram. We begin with 
the analysis of the analytic structure of the two loop integral. First, it is 
superficially divergent so one has
\[
\Pi_{\nu\rho}(\Delta,\epsilon) ~=~ \eta_{\nu\rho}\frac{R(\epsilon)}{\Delta} ~+~ 
\mbox{regular terms} ~.
\]
Second, for the same reason as we discussed for the previous graph 
$\Pi_{\nu\rho}(\Delta,\epsilon)$~$\to$~$0$ as $\epsilon$~$\to$~$0$. Third, the 
residue at the $\Delta$ pole is independent of $\epsilon$. Indeed, since the 
divergence is logarithmic the residue does not depend on the extermal momenta. 
Then, choosing a new route for the external momenta flow, for example along the
vertical line, one can see that all the dependence on $\epsilon$ disappears.
Since $\Pi_{\nu\rho}(\Delta,0)$~$=$~$0$, we conclude that $R$~$=$~$0$ and 
therefore the diagram is finite. Moreover, it is not hard to check that the 
terms proportional to $\eta_{\nu\rho}$ are even in $\epsilon$, and therefore we
have 
\begin{equation}
\Pi_{\nu\rho}(\Delta,\epsilon) ~=~ \left[ \epsilon^2 \eta_{\nu\rho} A ~+~ 
\frac{p_{\nu}p_\rho}{p^2} \left( \epsilon D +\epsilon^2 B +\epsilon \Delta C
\right) \right] ~.
\end{equation}  
For $\epsilon$~$=$~$\Delta$  the diagram with contracted indices can be 
reduced, in coordinate space, to the sum of chain integrals and one with a 
unique vertex of the Gross-Neveu type. Next, it can be checked that 
$g^{\nu\rho}\Pi_{\nu\rho}(\Delta,\Delta)$ $=$~$0$ which results in the 
following constraints on the coefficients
\[
D ~=~0 ~,  \ \ \ \ 2\mu A ~+~ B ~+~ C ~=~ 0 ~.
\]  
We recall that due to the presence of the transverse projector in 
(\ref{rightd}) we are only interested in the coefficient $A$. However, it 
is easier to calculate coefficients $B$ and $C$ and find $A$ from the above 
constraint. To determine $C$ we repeat the technique used for the evaluation of
the previous diagram. In other words we set $\epsilon$~$\to$~$\epsilon_{1(2)}$ 
for the upper and lower lines, respectively. Then, $C$ is equal to the 
coefficient of the $\epsilon\Delta$ term in the sum of the two diagrams 
$\Pi_{\nu\rho}(\Delta,\epsilon_1=0,\epsilon_2=\epsilon)$ 
$+$~$\Pi_{\nu\rho}(\Delta,\epsilon_1=\epsilon,\epsilon_2=0)$, which are easy to
integrate. To deduce $B$ we first transform the master diagram into 
coordinate space, giving 
\[
\frac{1}{(x^2)^{\mu-2\Delta}}{\tilde \Pi}_{\nu\rho}(x,\Delta,\epsilon) ~.
\]
It is easy to check that for $\Delta$~$=$~$0$, only the term proportional to
$B$ survives in ${\tilde \Pi}$ with ${\tilde \Pi}_{\nu\rho}(x,\Delta,\epsilon)$ 
$\sim$~$B (\eta_{\nu\rho}-2\mu\, x_\nu x_\rho/x^2)$~$+$~$O(\Delta)$. Next, we 
use tetrahedral symmetry whereby we add the new line, 
$x_\nu x_\rho/(x^2)^{1+\kappa}$, to the diagram to obtain a vacuum graph. The 
integration over $x$ yields the pole in $\kappa$ which is independent of which 
line we place the regularizing parameter, $\kappa$, \cite{Isaev}. Thus we can 
place the regulator on the vertical line and change the order of integration 
where we integrate over the upper vertex last. After some algebra the new 
diagram can be reduced to chains which allows us to determine the coefficient 
$B$, and hence the coefficient $A$ which we are interested in. 

Clearly, the evaluation of the three loop contribution to the gluon propagator
is a tedious exercise. To ensure that we have determined it correctly, aside 
from the checks we will discuss later, we have also undertaken to calculate it 
by another method. As this is equally as long an exercise we will briefly 
summarize the main steps. It is based on the original method of subtractions of
\cite{VPH} but differs from the one outlined above in that the original 
Feynman diagram is broken up into a sum of scalar integrals by taking spinor 
traces and the relevant Lorentz projections. Although this results in a large 
set of integrals the majority of them are in fact reducible to simple chain 
integrals or graphs involving the two loop self energy master diagram. The 
latter can readily be evaluated by uniqueness methods. The remaining three loop
graphs fall into two classes. Either they are divergent, and therefore their 
simple pole and finite part with respect to $\Delta$ need to be determined, or 
they are fully $\Delta$-finite. In the former case we were able to apply 
integration by parts and related methods to again reduce them to integrals 
which were chains, master diagrams or additionally three loop integrals which 
were evaluated by subtraction methods to the finite part. 
\begin{figure}[ht] 
\epsfig{file=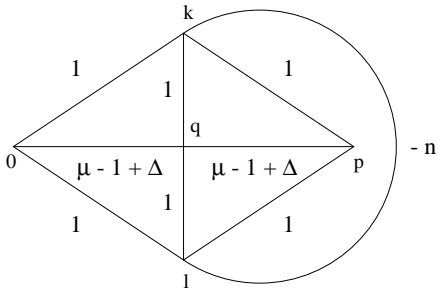,height=4cm} 
\vspace{0.5cm} 
\caption{Finite three loop diagrams contributing to gluon self-energy.} 
\label{glthreeloopfin} 
\end{figure}
The purely $\Delta$-finite integrals, in fact, represented the most difficult
to determine and were of the form given in figure \ref{glthreeloopfin} where
$n$~$=$~$0$, $1$, $2$ or $3$ and $k$, $l$ and $q$ are the loop momenta with $p$
the external momentum. In this notation the negative integer exponent 
corresponds to a numerator factor in the integrand, $((k-l)^2)^n$. The 
$n$~$=$~$0$ case had been treated previously in \cite{VPH}. To determine the 
other graphs we considered a related diagram which involved fermion lines 
constructed in such a way that when the $n$ traces were computed the original 
scalar diagram of figure \ref{glthreeloopfin} emerged. The additional integrals 
which accompanied this were again either chains or two loop self energy graphs,
aside of course from the new fermionic graph. This was evaluated by applying a 
Fourier transform to map it to coordinate space before taking the spinor 
traces. However, to avoid intermediate singularities in this step which arise 
from anti-uniqueness we introduced a temporary analytic regularization, 
$\delta$, in the exponents of the propagators. The advantage of transforming to 
coordinate space is that the resulting scalar integrals all have a vertex where
the scalar uniqueness relation can be applied to leave chain integrals. For 
completeness we note that the values of the integral of figure 
\ref{glthreeloopfin} are  
\begin{eqnarray} 
&& -~ \frac{(\mu-1)a^4(1) a^2(2\mu-2)}{2\Gamma(\mu)} \left[
\Phi(\mu) + \Psi^2(\mu) ~+~ \frac{a^3(1)a(2\mu-3)}{(2\mu-3)(\mu-2)} \right. 
\nonumber \\  
&& \left. ~~+~ \frac{(4\mu-7)(3\mu-4)\Psi(\mu)}{(2\mu-3)(\mu-2)(\mu-1)} ~+~ 
\frac{(13\mu^2-35\mu+23)}{(2\mu-3)(\mu-1)^2(\mu-2)} \right] 
\end{eqnarray}  
for $n$~$=$~$2$ and  
\begin{eqnarray} 
&& \frac{a^4(1) a^2(2\mu-2)}{2(2\mu-1)\Gamma(\mu)} \left[ \mu(\mu-1)
\left( \Phi(\mu) + \Psi^2(\mu) \right) ~-~ \frac{\mu(\mu-1)a^3(1)a(2\mu-3)} 
{(2\mu-1)(2\mu-3)(\mu-2)} \right. \nonumber \\  
&& \left. ~~~~~~~~~~~~~~~~~~~~~~+~ \frac{3(12\mu^4-48\mu^3+63\mu^2-30\mu+4) 
\Psi(\mu)}{(2\mu-1)(2\mu-3)(\mu-2)} \right. \nonumber \\ 
&& \left. ~~~~~~~~~~~~~~~~~~~~~~+~ \frac{(58\mu^5-232\mu^4+339\mu^3 
-230\mu^2+78\mu-12)}{(2\mu-1)(2\mu-3)(\mu-1)(\mu-2)\mu} \right] 
\end{eqnarray}  
for $n$ $=$ $3$. 
Including these with the other values we have verified that the same value for 
the three loop gluon self energy is obtained in the Landau gauge.  

This completes our detailed discussion of the evaluation of the three loop 
self energy diagrams and we now collect all the contributions and record the 
final values for all the relevant Feynman diagrams. First, we consider the 
gluon polarization tensor. To fix normalization we write down the expressions 
for the propagators in momentum space
\[ 
G_\psi ~=~ \frac{i\pslash}{p^2} ~, \ \ \ G_{gh} ~=~ \frac{1}{p^2} ~, \ \ \ 
G_{\nu\rho} ~=~ \frac{G\,M^{2\Delta}}{n (p^2)^{\mu-1+\Delta}} 
\left[\eta_{\nu\rho}- (1-\xi)\frac{p_\nu\, p_\rho}{p^2}\right] ~,
\] 
where the amplitude $G$ in the gluon propagator reads 
\begin{equation}
G ~=~ \frac{(4\pi)^\mu\Gamma(2\mu)}{2\Gamma^2(\mu)\Gamma(2-\mu)} ~.
\end{equation}
Further, the contribution from the $i$-th graph of figure \ref{fig2} to the 
polarization tensor $\Pi_{\mu\nu}$ can be written in momentum space as 
\begin{equation}
\Pi_{\mu\nu}^{ab,i}(p) ~=~ \frac{\delta^{ab}}{(p^2)^{1-\mu+n_i\Delta}}
\left (A^i P_{\mu\nu}^\perp+B^i P_{\mu\nu}^\parallel\right) ~, \ \ \ \ 
\Pi_{\mu\nu} ~=~ \sum_i\Pi_{\mu\nu}^i ~.
\label{selfenergy}
\end{equation}
Here $P^\perp$, $P^\parallel$ are the transverse and longitudinal projectors
and $n_i$ is the number of the gluon lines in the diagram. The calculation 
yields for the coefficients $B^i$
\begin{eqnarray}
&&B^a ~=~ -~ (C_F-C_A/2)\,R_1 ~, \ \ \ \ \ \ \ \ B^{b+c} ~=~ C_F\, R_1 ~, 
\nonumber\\
&&B^d ~=~ C_A\,R_2 ~, \ \ \ \  \ \ \ \ \ \ \ \ \ \  \ \ \ \ \ \ \ \ B^{e+f} ~=~
-~ C_A\,(R_1+R_2) ~, 
\end{eqnarray}
where
\begin{eqnarray}
R_1&=& \frac{(2\mu-1)(\mu-2)}{(4\pi)^\mu \mu(\mu-1)\Gamma(\mu)}\left[
1+\frac{\xi\mu}{(2\mu-1)(\mu-2)}
\right] ~, \nonumber \\
&& \nonumber \\
R_2 &=& -~ \frac{\Gamma^2(\mu)a(2\mu-1)}{2(4\pi)^\mu} ~.
\end{eqnarray}
As can be seen they are all finite and their sum vanishes, $\sum_i B^i$ $=$ 
$0$.

Instead of recording the values for $A^i$ we list the values for the 
coefficients $\Pi^i$, which are related to the former as
\[
A^i ~=~ \frac{\Pi^i-B^i}{(2\mu-1)} ~,
\]
and, as is easy to see, they are given by the trace of the corresponding 
contributions to the polarization tensor, $\Pi^i$ $=$ $\Pi^{i~\,\nu}_{~\nu}$. 
We find 
\begin{eqnarray}
\Pi^a&=&2\left ( C_F-\frac{C_A}{2}\right) R(\Delta)\times \nonumber \\
&&\nonumber \\
&&\times\left[
\frac{1}{\Delta}\left (1+\frac{\xi\mu}{(\mu-2)(2\mu-1)}\right )-
\frac{3\mu(\mu-1)}{(\mu-2)(2\mu-1)}\left( \Theta(\mu) - \frac{1}{(\mu-1)^2} 
\right) \right] ~, \\
&&\nonumber \\
\Pi^b&=&\Pi^c ~=~ -~ C_F\,R(\Delta)\,\left[\frac{1}{\Delta}
\left( 1+\frac{\xi\mu}{(\mu-2)(2\mu-1)}\right )
+\frac{2(\mu-1)}{\mu(\mu-2)} \right] ~, \\
&&\nonumber \\
\Pi^d&=& -~ \frac{C_A}{(4\pi)^\mu}\,\Gamma^2(\mu)\, a(2\mu-1) ~, \\
&&\nonumber \\
\Pi^{ef}&=&C_A\,\frac{R(2\Delta)}{2}\left(\frac{\mu-1}{\mu-2}\right)\left[
\frac{1}{\Delta}\left(1+\frac{\xi}{(\mu-1)}\right) \right.
-\frac{2\,\mu\,\Gamma^3(\mu)\,a(2\mu-1)}{(\mu-1)\,(2\mu-1)^2} \nonumber \\
&&\nonumber \\
&& +~ \frac{2\mu}{(2\mu-1)}\left(\Psi^2(\mu)+\Phi(\mu)-6\Theta(\mu)\right) ~-~ 
\frac{\mu(8\mu^{4}-76\mu^{3}+198\mu^2-193\mu+62)}{(2\mu-1)^{2}(2\mu-3)(\mu-1) 
(\mu-2)}\Psi(\mu) \nonumber \\
&&\nonumber \\
&& +~ \frac{2(8\mu^{7}-60\mu^{6}+206\mu^{5}-374\mu^{4}+346\mu^{3}-149\mu^{2} 
+34\mu- 6)}{(2\mu - 1)^{2}(2\mu - 3)(\mu - 1)^{2}(\mu - 2)\mu} \\
&&\nonumber \\
&&\left. -~ \frac{2\xi\Psi(\mu)}{(2\mu-1)} ~-~ 
\frac{\xi(10\mu^4-34\mu^3+27\mu^2+10\mu-12)}
{\mu(\mu-1)^2(\mu-2)(2\mu-1)(2\mu-3)} ~-~ \frac{\xi^2\mu} 
{2(\mu-1)^2(2\mu-1)} \right] ~. \nonumber
\end{eqnarray}
Here
\[
R(\Delta) ~=~ -~ \frac{(\mu-2)(2\mu-1)^2 a(\mu-1+\Delta) a(1-\mu)} 
{(4\pi)^\mu \mu(\mu-1+\Delta) a(1-\mu+\Delta)} ~,
\]
and the functions $\Psi(\mu)$, $\Phi(\mu)$ and $\Theta(\mu)$ are defined as
\begin{eqnarray}
\Psi(\mu) &=& \psi(2\mu-3) ~+~ \psi(3-\mu) ~-~ \psi(1) ~-~ \psi(\mu-1) ~, 
\nonumber\\
\Phi(\mu) &=& \psi^\prime(2\mu-3) ~-~ \psi^\prime(3-\mu) ~-~ 
\psi^\prime(\mu-1) ~+~ \psi^\prime(1) ~, \\
\Theta(\mu) &=& \psi^\prime(\mu-1) ~-~ \psi^\prime(1) ~, \nonumber
\end{eqnarray}
where $\psi(x)$ $=$ $(\ln{\Gamma(x)})^\prime$. Taking into 
account~(\ref{selfenergy}) the renormalized gluon   
propagator is
\begin{equation}
G_{\nu\rho}(p) ~=~ \frac{G}{n (p^2)^{\mu-1}} \left[
\left(1~+\frac1n\left[r~+~\gamma_A^{(1)}\,\ln\left({p^2}/{M^2}\right)\right]\,
\right) P^\perp_{\nu\rho}~+~\xi P^\parallel_{\nu\rho} \right] ~,
\end{equation}
where
\begin{eqnarray}
\gamma_A^{(1)}&=& -~ \sum_i n_i\Pi_{-1}^{(i)} ~=~ \frac{C_A \eta_0}{2(\mu-2)}
\left(1+\frac{\xi(\mu-1)}{(2\mu-1)}\right) ~, \\
&&\nonumber \\
r&=& \sum_i \Pi_0^{(i)} ~=~ \frac{\eta_0}{2(\mu-2)}\left[ C_F\left(
\frac{6\mu\,(\mu-1)}{(2\mu-1)}\left(\Theta(\mu)-\frac{1}{(\mu-1)^2}\right)+
\frac{4(\mu-1)}{\mu}\right) \right.\nonumber \\
&&\nonumber \\
&& +~ \frac{C_A}{(2\mu-1)}\left( \frac{}{} -~ \mu(\mu-1) \left( \Psi^2(\mu) 
+\Phi(\mu) - 3\Theta(\mu) \right) \right. \nonumber \\ 
&&\nonumber \\
&& \left. +~ \frac{(8\mu^5-92\mu^4+270\mu^3-301\mu^2+124\mu-12)\Psi(\mu)} 
{2(\mu-2)(2\mu-1)(2\mu-3)} \right. \nonumber \\
&&\nonumber \\
&& -~ \frac{(16\mu^7-120\mu^6+420\mu^5-776\mu^4+742\mu^3-349\mu^2+84\mu-12)} 
{2\mu(\mu-1)(\mu-2)(2\mu-1)(2\mu-3)}\nonumber \\
&&\nonumber \\
&&\left.\left. +~ \frac{\xi}{2}\left(\frac{(\mu^2+2\mu-2)}{\mu(\mu-1)}+
\frac{\xi\mu}{2(\mu-1)}\right)\right)\right] ~,
\end{eqnarray}
and the functions $\Pi_0^{(i)}$ and $\Pi_{-1}^{(i)}$ are defined as 
\begin{equation}
\Pi^{(i)} ~=~ \eta_0\, (2\mu-1) \, G^{-1} \left[ 
\frac{\Pi^{(i)}_{-1}}{\Delta}+\Pi^{(i)}_0\right] ~.
\end{equation}
Finally, $\eta_0$ is twice the anomalous quark dimension in Landau gauge which 
is recorded in (\ref{indices}),  
\begin{equation}
\eta_{0} ~=~ \frac{(\mu-2)(2\mu-1)\Gamma(2\mu)} 
{\Gamma^2(\mu)\Gamma(\mu+1)\Gamma(2-\mu)} ~.
\end{equation}
We note that the part of the polarization tensor corresponding to the QED 
contribution is independent of the gauge fixing parameter $\xi$ which is 
consistent with a general analysis. For example, see \cite{Zinn}. 

Next we turn to the quark propagator and record the values for the graphs of 
figure \ref{fig1}. First, for completeness we write down the value for 
the graph 

\centerline{\epsfxsize12.0cm\epsfbox[140 430  450 480]{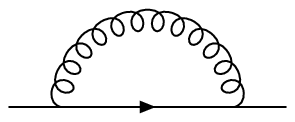}} 

\noindent
which contributes to the fermion propagator at $O(1/\Nf)$. Straightforward
calculations give  
\begin{equation}
i\pslash \left (\frac{4M^2}{p^2}
\right)^\Delta \> \frac{C_F \eta_0}{2\Delta}
\left [1+\frac{\xi\mu}{(2\mu-1)(\mu-2)}+\frac{2(\mu-1)\Delta}{\mu(\mu-1)} 
+ O(\Delta^2) \right]
\end{equation}
for the above graph which results in the following anomalous quark dimension at 
$O(1/\Nf)$
\begin{equation}
\gamma_\psi^{(1)} ~=~ \frac{C_F\eta_0}{2} 
\left( 1 + \frac{\xi\mu}{(2\mu-1)(\mu-2)} \right)
\end{equation}
and the renormalized propagator
\begin{equation}
G_\psi(p) ~=~ \frac{i\pslash}{p^2}\left[ 1+\frac{C_F}{n}\left( -~  
\frac{(\mu-1)\eta_0}{\mu(\mu-2)} ~+~ \gamma_q^{(1)}\ln{\left(p^2/M^2\right)} 
\right) \right] ~.
\label{gluon_rp}
\end{equation}
Here and elsewhere we use the notation $A$ $=$ $\sum_k A^{(k)}/n^k$.

Below we list the ${\cal KR}^\prime$ values of the graphs of figure \ref{fig1},
which are necessary to determine the quark anomalous dimension at $O(1/\Nf^2)$.
The value of each graph after application of the ${\cal KR}^\prime$ operation 
will be denoted by the respective capital letter. So, for example, applying
${\cal KR}^\prime$ to graph (b) of \ref{fig1} gives $i\pslash\, B$. In addition
$A^{(i)}$ corresponds to the value of the $i$-th subgraph of figure \ref{fig2}.
We find  
\begin{eqnarray}
A^{(i)}&=&C_F\,\eta_0\left [
\frac{(n_i-2)}{2 n_i \Delta^2}\Pi^{(i)}_{-1}+\frac{1}{2n_i\Delta}\left(
\Pi_0^{(i)}-\frac{B^{(i)}}{(2\mu-1)}\left(1-\frac{\xi\mu}{(\mu-2)} 
\right)\right) \right] ~, \\
&&\nonumber \\
B&=&\frac{C_F^2\eta_0^2}{8}\,\left(1+\frac{\xi\mu}{(\mu-2)(2\mu-1)}\right)^2
\left[\frac{1}{\Delta^2}-\frac{1}{\Delta\,\mu}\right] ~,\\
C&=& -~ C_F\,\left(C_F-\frac{C_A}{2}\right)\frac{\eta_0^2}{4} \nonumber\\
&& \times \left [ \left(\frac{1}{\Delta^2}-\frac{1}{2\Delta\mu}\right )
\left(1+\frac{\xi\mu}{(\mu-2)(2\mu-1)}\right)^2 + 
\frac{2(\mu-1)(\mu^2-\mu-1)}{\mu(\mu-2)^2(2\mu-1)\Delta}\right] ~, \\
&&\nonumber \\
D+E&=& -~ \frac{C_A\eta_0^2}{12}\left[
\frac{(\mu-1+\xi)[(\mu-2)(2\mu-1)+\xi\mu]}{(\mu-2)^2(2\mu-1)\Delta^2} 
\right.\nonumber \\
&&\nonumber \\
&&+~ \frac{3\mu(\mu-1)}{(\mu-2)(2\mu-1)\Delta}\Theta(\mu) ~-~ 
\frac{4\mu^4-10\mu^3+9\mu^2+4\mu-2}{2\mu(\mu-1)(\mu-2)(2\mu-1)\Delta} \\
&&\nonumber \\
&&\left. +~ \frac{\xi\,(\mu^2-4\mu+2)}{\mu(\mu-2)^2(\mu-1) 
(2\mu-1)\Delta} \left(1+\frac{\xi\mu}{2} \right)\right] ~.
\end{eqnarray}
According to (\ref{sf}) the anomalous quark dimension at $O(1/\Nf^2)$ is given
by the sum of the residues at simple poles in $\Delta$ multiplied by the number
of gluon lines in the diagrams. The final result for $\eta_2$ will be given in
section 7. 

\sect{Schwinger Dyson approach.} 

In this section we present the Schwinger Dyson, (SD), formalism to determine 
the exponent $\eta_2$. There are various motivations for this. One is that in 
the context of the newer approach of the previous sections it will allow an 
interested reader to compare and contrast the technology involved. Though in
fact it will not represent a significant amount of extra work given that the 
difficult Feynman integrals have now been evaluated. Second, it provides 
another check on our calculations as it is clear these are quite technical. 
Third, we would like to discuss some distinctions in the treatment of the
theories with abelian and nonabelian gauge symmetries in the SD approach. Our 
first comment concerns the issue of choosing the gauge fixing parameter, $\xi$,
entering the gluon propagator. To appreciate the subtlety of this choice we 
recall that in the original construction of Vasil'ev et al for scale invariant 
theories, such as $\phi^4$ theory, \cite{VPH}, one omits at the outset the 
contribution of the bare propagator in the SD equation, which is less singular 
in the infra-red region compared to the loop corrections. This converts the SD 
equations into self-consistency equations which determine the exponents of the 
propagators. In respect of the discussion of the previous sections concerning 
QCD and the NATM, scale invariance is only present in two gauges. These are the
Landau gauge, $\xi$ $=$ $0$, and that given by $\xi$ $=$ $\xi_A$ in 
(\ref{xi_A}). Therefore, strictly speaking, our subsequent SD analysis will 
only be valid for these particular gauges. By contrast in an abelian gauge 
theory such as QED the gauge propagator is scale invariant in any gauge in the 
$1/\Nf$ expansion. To understand how the choice of these two gauges arises from
demanding scale invariance one can repeat the arguments given in section 2. 
Indeed, on general grounds the form of the full gluon propagator in the 
massless limit will take the form  
\begin{equation} 
\tilde{A}_{\mu\nu}(k) ~=~ \frac{\tilde{B}^\perp}{(k^2)^{\mu-\beta}} \left[ 
\eta_{\mu\nu} ~-~ \frac{k_\mu k_\nu}{k^2} \right] ~+~ 
\tilde{B}^\parallel \xi \frac{k_\mu k_\nu}{(k^2)^\mu} ~.  
\end{equation} 
Clearly to have a scale invariant gluon propagator the exponents of the 
momentum prefactor in the transverse and longitudinal pieces both ought to be
the same. This can come about in two ways subject to the physical restriction
of retaining the transverse part. One case is to have no longitudinal part 
which gives rise to the Landau gauge, $\xi$ $=$ $0$. The other is to match the 
powers of the exponents in each term. This requires the gluon to have zero 
anomalous dimension, $\eta$ $+$ $\chi$ $=$ $0$. Since this exponent is gauge 
dependent then one can in principle solve this equation order by order in large
$\Nf$ perturbation theory to determine the explicit value of $\xi$. It is this 
solution which we denote by $\xi_A$. It is worth noting that in QED, 
\cite{JG91,JG94}, the photon anomalous dimension vanishes in all gauges by
virtue of the QED Ward identity and therefore the ansatz for the photon 
propagator was automatically scale invariant. Alternatively another point of 
view can be taken on this problem. One can solve the SD equations in a 
nonabelian gauge theory taking a scale invariant ansatz for the gluon
propagator with an arbitrary gauge fixing parameter. This parameter is then 
considered as an input parameter which is to be determined. Therefore, it is 
not hard to understand that solving the SD equations one will find that the 
latter are only consistent for two particular values of $\xi$. These will be 
$\xi$ $=$ $0$ and $\xi$ $=$ $\xi_A$.

In light of these observations we now proceed with the SD calculation but 
restrict ourselves to the Landau gauge. Therefore, the ansatz for the gluon 
propagator in this calculation is,  
\begin{equation} 
\tilde{A}_{\mu\nu}(k) ~=~ \frac{\tilde{B}}{(k^2)^{\mu-\beta}} \left[ 
\eta_{\mu\nu} ~-~ \frac{k_\mu k_\nu}{k^2} \right] 
\label{Amomprop} 
\end{equation} 
in momentum space. The gluon propagator exponent is given by
\begin{equation} 
\beta ~=~ 1 ~-~ \eta ~-~ \chi 
\end{equation} 
where $\chi$ is the quark gluon vertex anomalous dimension. For the quark and
ghost fields we define their momentum space propagators as  
\begin{equation} 
\tilde{\psi}(k) ~=~ \frac{\tilde{A}\kslash}{(k^2)^{\mu-\alpha}} ~~~,~~~ 
\tilde{c}(k) ~=~ \frac{\tilde{C}}{(k^2)^{\mu-\gamma}} 
\label{qmomprop} 
\end{equation} 
where $\tilde{A}$, $\tilde{B}$ and $\tilde{C}$ are the momentum independent 
amplitudes. The analogous coordinate space amplitudes will be $A$, $B$ and $C$ 
respectively. In the SD critical point method, \cite{VPH}, one determines the 
unknown exponent, $\eta$, by representing the two point SD equations at the 
appropriate order which is $O(1/\Nf^2)$. The representation of the equations 
form a set of algebraic equations with various unknowns one of which is $\eta$ 
whilst the others are combinations of the amplitudes $\tilde{A}$, $\tilde{B}$ 
and $\tilde{C}$. Eliminating the latter variables allows one to deduce $\eta$.  
\begin{figure}[ht] 
\epsfig{file=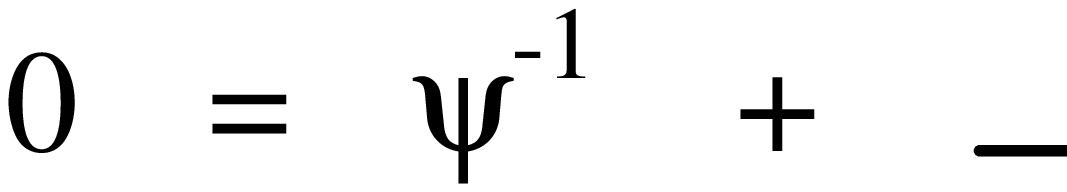,width=14cm} 
\vspace{0.5cm} 
\caption{Quark Schwinger Dyson equation to $O(1/\Nf^2)$.} 
\label{qSDfig} 
\end{figure}

To illustrate these points we focus initially on the quark equation as there
are additional features which need to be considered in the treatment of the 
gluon equation which will be discussed later. If we denote by $\Sigma_1$ and
$\Sigma_2$ the values of the respective two and three loop graphs of figure 
\ref{qSDfig} in coordinate space then the SD equation at $O(1/\Nf^2)$ can be 
represented algebraically by 
\begin{eqnarray} 
0 &=& r(\alpha-1) ~+~ \Sigma_0 z m^2 (x^2)^{\chi+\Delta} ~+~ C_F (C_F - C_A/2) 
(x^2)^{2\chi+2\Delta} z^2 \Sigma_1 \nonumber \\ 
&& +~ C_F C_A (x^2)^{3\chi+3\Delta} T_F \Nf z^3 \Sigma_2 
\label{qSD} 
\end{eqnarray}  
where 
\begin{equation} 
\Sigma_0 ~=~ 2\left[ \frac{(\beta-\Delta)}{(2\mu-2\beta-1+2\Delta)} ~-~ (\mu-1) 
\right] ~, 
\end{equation} 
corresponds to the first graph of figure \ref{qSDfig} and $z$ $=$ $A^2B$. The 
analytic regularization has been included by shifting the vertex anomalous 
dimension from $\chi$ to $\chi$ $+$ $\Delta$. As in the original method of 
\cite{VPH} we have chosen the coordinate space representation of the Schwinger 
Dyson equation. To relate to the momentum space version one applies the usual 
Fourier transform 
\begin{equation} 
\frac{1}{(x^2)^\alpha} ~=~ \frac{a(\alpha)}{2^{2\alpha}\pi^\mu} \int_k 
\frac{e^{ikx}}{(k^2)^{\mu-\alpha}} 
\end{equation}   
to the version of (\ref{qSD}) prior to the powers of $x^2$ being cancelled. The
coordinate space propagators can be determined from (\ref{Amomprop}) and 
(\ref{qmomprop}) and are taken to be  
\begin{eqnarray} 
\psi(x) &=& \frac{A\xslash}{(x^2)^\alpha} ~~~,~~~ 
A_{\mu\nu}(x) ~=~ \frac{B}{(x^2)^\beta} \left[ \eta_{\mu\nu} ~+~ 
\frac{2\beta}{(2\mu-2\beta-1)} \frac{x_\mu
x_\nu}{x^2} 
\right] 
\nonumber \\ 
c(x) &=& \frac{C}{(x^2)^\gamma} ~. 
\end{eqnarray} 
The inverse quark propagator, which is the origin of the first term of 
(\ref{qSD}), is given by 
\begin{equation} 
\psi^{-1}(x) ~=~ \frac{r(\alpha-1)\xslash}{A(x^2)^{2\mu-2\alpha+1}}  
\end{equation} 
where 
\begin{equation} 
r(\alpha) ~=~ \frac{\alpha a(\alpha-\mu)}{\pi^{2\mu}(\mu-\alpha)a(\alpha)} ~.  
\end{equation} 
In (\ref{qSD}) the quantity $m$ corresponds to the vertex renormalization 
constant which will absorb the divergences arising in the two and three loop
corrections. To make this more explicit we define their $\Delta$-dependence as  
\begin{equation} 
\Sigma_i ~=~ \frac{K_i}{\Delta} ~+~ \Sigma^\prime_i 
\end{equation} 
where the $O(\Delta)$ terms are not important at this order in $1/\Nf$. 
Therefore the quark Dyson equation is rendered finite by defining the 
counterterm formally as 
\begin{equation} 
m_1 ~=~ \frac{(2\mu-3)z_1}{4(2\mu-1)(\mu-2)T_F} \left[ (C_F - C_A/2) K_1 ~+~ 
C_A z_1 K_2 \right] 
\label{mq} 
\end{equation}  
where we have set 
\begin{equation} 
m ~=~ 1 ~+~ \frac{m_1}{\Nf\Delta} ~+~ O \left( \frac{1}{\Nf^2} \right) ~~~,~~~ 
z ~=~ \frac{z_1}{\Nf} ~+~ \frac{z_2}{\Nf^2} ~+~ O \left( \frac{1}{\Nf^3} 
\right) ~.  
\end{equation}  
However, this leaves terms in the SD equation involving $\ln x^2$ which would 
otherwise spoil the simple scaling behaviour at the fixed point. To avoid this 
one defines the vertex anomalous dimension to be 
\begin{equation} 
\chi_1 ~=~ \frac{(2\mu-3)z_1}{2(2\mu-1)(\mu-2)T_F} 
\left[ (C_F - C_A/2) K_1 ~+~ 2 C_A z_1 K_2 \right] ~.  
\label{chiq} 
\end{equation}  
With these definitions the finite quark Dyson equation to $O(1/\Nf^2)$ is
\begin{eqnarray} 
0 &=& r(\alpha-1) ~+~ C_F \left[ \frac{\Sigma_{00}z_1}{\Nf} ~+~ 
\frac{(\Sigma_{00}z_2 ~+~ \Sigma_{02}z_1 + 2m_1 \Sigma_{01}z_1)}{\Nf^2}
\right] 
\nonumber \\ 
&& +~ C_F (C_F - C_A/2) \frac{z_1^2}{\Nf^2} \Sigma^\prime_1 ~+~ 
C_F C_A \frac{z_1^3}{\Nf^2} \Sigma^\prime_2  
\label{qSDfin} 
\end{eqnarray}  
where the first term contains $\eta_1$ and $\eta_2$ and we have expanded 
$\Sigma_0$ as  
\begin{equation} 
\Sigma_0 ~=~ \Sigma_{00} ~+~ \Delta \Sigma_{01} ~+~ \frac{(\Sigma_{02} + \Delta
\Sigma_{03})}{\Nf} ~+~ O \left( \frac{1}{\Nf^2} ; \Delta^2 \right) ~.  
\end{equation}  
Therefore, from (\ref{qSDfin}) we can determine a relation for $\eta_2$ in 
terms of $z_2$ and the finite correction graphs, as 
\begin{equation} 
\eta_2 ~=~ \frac{z_2\eta_1}{z_1} ~+~ \frac{\eta_1^2}{2\mu} ~+~ \left[ 
\Sigma_{02} + 2m_1\Sigma_{01} + (C_F - C_A/2)z_1\Sigma^\prime_1 + C_A z_1^2 
\Sigma^\prime_2 \right] \frac{\eta_1}{\Sigma_{00}} 
\label{eta2defn} 
\end{equation}  
where $\eta$ $=$ $\sum_{i=1}^\infty \eta_i/\Nf^i$ and $\eta_1$ $=$ 
$\gamma^{(1)}/(2T_F)$ in earlier notation.  

\begin{figure}[ht] 
\epsfig{file=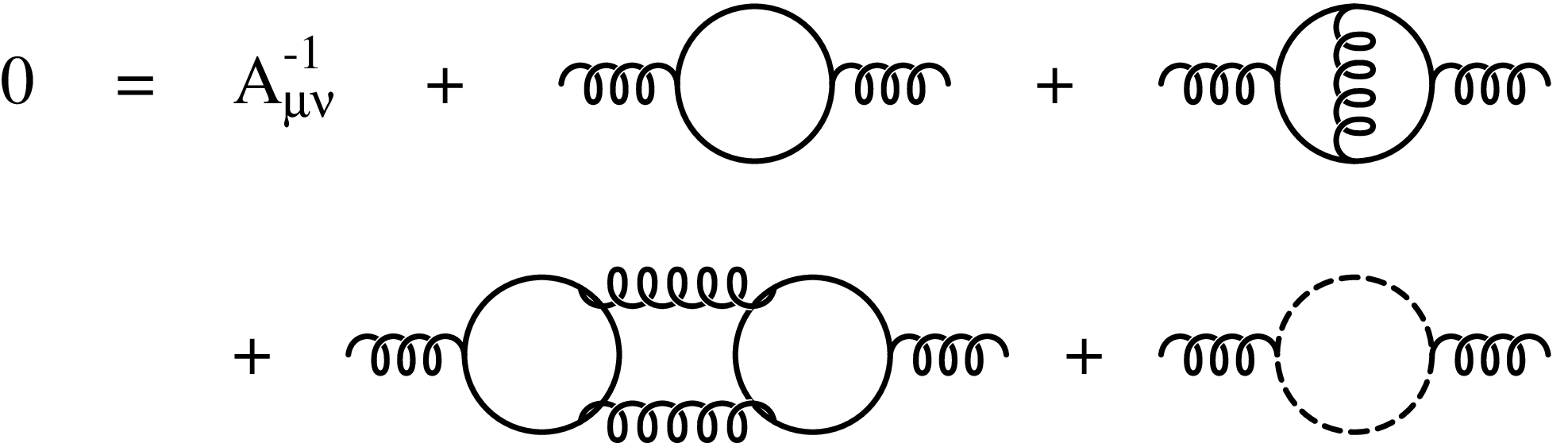,width=12cm} 
\vspace{0.5cm} 
\caption{Gluon Schwinger Dyson equation to $O(1/\Nf^2)$.} 
\label{gSDfig} 
\end{figure}

For the gluon SD equation there is the added complication of dealing with the 
transverse and longitudinal components of the equation. Whilst this has been 
dealt with for QED in \cite{JG94} we will recall the important features. In 
coordinate space the inverse gluon propagator is, \cite{VPH81,JG94},  
\begin{equation} 
A^{-1}_{\mu\nu}(x) ~=~ \frac{m(\beta)}{B(x^2)^{2\mu-\beta}} \left[ 
\eta_{\mu\nu} ~+~ \frac{2(2\mu-\beta)}{(2\beta-2\mu-1)} 
\frac{x_\mu x_\nu}{x^2} \right] 
\end{equation} 
where 
\begin{equation} 
m(\beta) ~=~ \frac{(2\mu-2\beta+1)(2\mu-2\beta-1) 
a(\beta-\mu)} {4\pi^{2\mu}(\mu-\beta)^2 a(\beta)} ~.  
\end{equation} 
We recall, \cite{VPH81,JG94}, that this is deduced by inverting the gauge 
dependent momentum space gluon propagator on the physical transverse subspace 
before Fourier transforming back to coordinate space. Therefore, we can 
formally represent the coordinate gluon SD equation as 
\begin{eqnarray} 
0 &=& m(\beta-\Delta) \left[ \eta_{\mu\nu} 
+ \frac{2(2\mu-\beta+\Delta)}{(2\beta-2\mu-1-2\Delta)} \frac{x_\mu x_\nu}{x^2} 
\right] ~-~ 4z T_F\Nf m^2 (x^2)^{\chi+\Delta} \left[ \eta_{\mu\nu} 
- \frac{2x_\mu x_\nu}{x^2} \right] \nonumber \\ 
&& -~ (C_F - C_A/2) T_F\Nf z^2 (x^2)^{2\chi+2\Delta} \left[ 
\eta_{\mu\nu} \Pi_1 + \frac{x_\mu x_\nu}{x^2} \Xi_1 \right] \nonumber \\
&& +~ C_A T_F^2\Nf^2 z^3 (x^2)^{3\chi+3\Delta} \left[ \eta_{\mu\nu} \Pi_2 
+ \frac{x_\mu x_\nu}{x^2} \Xi_2 \right] \nonumber \\ 
&& -~ \frac{C_A\gamma^2(2\mu-2\beta-1) a^2(\mu-\gamma) 
(x^2)^{\chi_c+\Delta}\tilde{y}}{2(\mu-\beta)a(\beta)} \frac{x^\mu x^\nu}{x^2}  
\label{gSD} 
\end{eqnarray} 
where $\Pi_1$ and $\Xi_1$ relate to the coordinate space value of the two loop 
diagram of figure \ref{gSDfig} and $\Pi_2$ and $\Xi_2$ correspond to that of 
the three loop one. The final term of (\ref{gSD}) is the ghost contribution 
from the ghost graph of figure \ref{gSDfig}. However, its treatment requires 
special care. Although we have constructed the gluon equation in coordinate 
space, the ghost contribution is determined from its evaluation in momentum 
space since the ghost couples via a derivative to the gluon. This explains the 
appearance of the amplitude combination $\tilde{y}$ $=$ $\tilde{C}^2 \tilde{B}$
in (\ref{gSD}). In particular the ghost graph of figure \ref{gSDfig} is 
\begin{equation} 
-~ C_A \nu(\mu-\gamma,\mu-\gamma,2\gamma+1) \frac{\gamma^2\tilde{C}^2} 
{2(2\gamma+1)(p^2)^{\mu-2\gamma-1}} \left[ \eta^{\mu\nu} ~+~ 
2(2\gamma-\mu+1) \frac{p^\mu p^\nu}{p^2} \right] ~.  
\label{ghostmom} 
\end{equation}  
\begin{figure}[hb] 
\epsfig{file=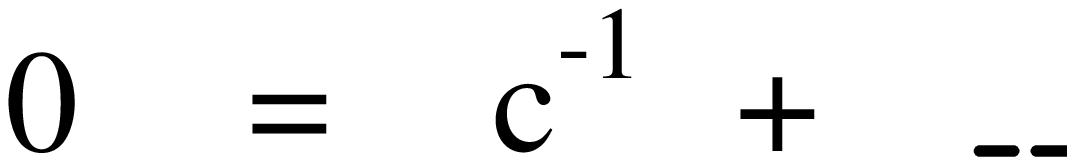,width=6cm} 
\vspace{0.5cm} 
\caption{Ghost Schwinger Dyson equation at $O(1/\Nf)$.} 
\label{ghSDfig} 
\end{figure}

Although the ghost was treated in large $\Nf$ in the Landau gauge in 
\cite{JG93} where its anomalous dimension and vertex anomalous dimension were
computed at $O(1/\Nf)$, the value of $\tilde{y}_1$ was not recorded. To 
determine it we consider the ghost SD equation of figure \ref{ghSDfig}. In 
momentum space it becomes 
\begin{equation} 
0 ~=~ 1 ~+~ (2\mu-1) \tilde{y} C_A 
\frac{\nu(\mu-\gamma-1,\mu-\beta,\beta+\gamma)}{2(\mu-\gamma-1)(\beta+\gamma)
(\mu-\beta)} 
\end{equation} 
whence 
\begin{equation} 
\tilde{y}_1 ~=~ -~ \frac{\mu\Gamma(\mu) \eta_{c,1}}{(2\mu-1)C_A} 
\label{ytilde} 
\end{equation} 
where $\eta_c$ $=$ $\sum_{i=1}^\infty \eta_{c,i}/\Nf^i$. Alternatively using 
the Slavnov-Taylor identity in the Landau gauge we have  
\begin{equation} 
\tilde{y}_1 ~=~ \frac{\Gamma(\mu+1)\eta_0}{2(\mu-2)(2\mu-1)T_F} ~.  
\end{equation}  
We note that in (\ref{gSD}) the $x$-dependence of the final term involves 
$\chi_c$. This emerges after using the Slavnov-Taylor identity in rationalising
powers of $x^2$ in the initial representation of the SD equation. 

Whilst there appears to be two components of (\ref{gSD}) to analyse it turns 
out that only one is important. This is the transverse component when expressed
in momentum space using the Fourier transform. Although this may appear to
neglect information contained in the longitudinal part of the equation with
respect to momentum space it turns out in fact that the sum of contributions to
the longitudinal projection of (\ref{gSD}) in momentum space from the various 
graphs in arbitrary gauge is actually zero. The cancellation in the $C_F$ 
sector was discussed in \cite{JG94}. For the $C_A$ sector the contributions 
from the longitudinal sectors of the last three graphs of figure \ref{gSDfig} 
sum to zero which can be verified from the explicit values of the graphs. This 
is also true for non-zero $\xi$. 

Therefore, returning to (\ref{gSD}) and restoring the common factor of 
$1/(x^2)^{2\mu-\beta+\Delta}$ in each term we transform the equation to 
momentum and retain only the transverse part. Inverting this component to 
coordinate space yields the relevant part of the gluon SD equation for 
determining the critical exponents,   
\begin{eqnarray}  
0 &=& \frac{2(\mu-\beta+\Delta)m(\beta-\Delta)} 
{(2\mu-2\beta+1+2\Delta)} ~-~ \frac{8(\alpha-1) z T_F\Nf m^2 
(x^2)^{\chi+\Delta}}{(2\alpha-1)} \nonumber \\ 
&& -~ (C_F - C_A/2) T_F\Nf z^2 (x^2)^{2\chi+2\Delta} \left[ \Pi_1 + 
\frac{\Xi_1}{2(2\alpha-1-\chi-\Delta)} \right] \nonumber \\
&& +~ C_A T_F^2 \Nf^2 z^3 (x^2)^{3\chi+3\Delta} \left[ \Pi_2 
+ \frac{\Xi_2}{2(2\alpha-1-2\chi-2\Delta)} \right] \nonumber \\
&& -~ \frac{C_A\gamma^2(2\mu-2\beta-1) a^2(\mu-\gamma) 
(x^2)^{\chi_c+\Delta}\tilde{y}}{4(2\gamma+1)(\mu-\beta)a(\beta)} 
\end{eqnarray} 
which now requires renormalization. This is performed in the same way that 
(\ref{qSD}) was rendered finite. Repeating the procedure here and ensuring
that there is no anomalous scaling behaviour we find, 
\begin{eqnarray}  
0 &=& \frac{2(\mu-\beta)m(\beta)}{(2\mu-2\beta+1)} ~-~ 
\frac{8(\alpha-1) z T_F\Nf}{(2\alpha-1)} \nonumber \\ 
&& -~ (C_F - C_A/2) T_F\Nf z^2 \left[ \Pi^\prime_1 
+ \frac{\Xi^\prime_1}{2(2\alpha-1)} + \frac{X_1}{2(2\alpha-1)^2} \right] 
\nonumber \\
&& +~ C_A T_F^2 \Nf^2 z^3 \left[ \Pi^\prime_2 
+ \frac{\Xi^\prime_2}{2(2\alpha-1)} + \frac{X_2}{(2\alpha-1)^2} \right] ~-~ 
\frac{C_A\gamma^2(2\mu-2\beta-1) a^2(\mu-\gamma) \tilde{y}} 
{4(2\gamma+1)(\mu-\beta)a(\beta)} \nonumber \\  
\label{gSDfin}  
\end{eqnarray} 
where we have defined 
\begin{equation} 
\Pi_i ~=~ \frac{P_i}{\Delta} ~+~ \Pi^\prime_i ~~~,~~~ 
\Xi_i ~=~ \frac{X_i}{\Delta} ~+~ \Xi^\prime_i 
\end{equation} 
for $i$ $=$ $1$ and $2$, and 
\begin{eqnarray} 
m_1 &=& -~ \frac{(2\mu-1)z_1}{16(\mu-1)} \left[ \left(C_F - \frac{C_A}{2} 
\right) \left( P_1 + \frac{X_1}{2(2\mu-1)} \right) ~-~ C_A z_1 \left( P_2 
+ \frac{X_2}{2(2\mu-1)} \right) \right] \nonumber \\  
\chi_1 &=& -~ \frac{(2\mu-1)z_1}{8(\mu-1)} \left[ \left(C_F - \frac{C_A}{2} 
\right) \left( P_1 + \frac{X_1}{2(2\mu-1)} \right) ~-~ 2 C_A z_1 \left( P_2 
+ \frac{X_2}{2(2\mu-1)} \right) \right] ~. \nonumber \\   
\label{mg} 
\end{eqnarray} 
Although it may seem that (\ref{mq}) and (\ref{chiq}) will give different 
values from (\ref{mg}), when the explicit values for the residues of the simple
poles in $\Delta$ are substituted, they are equivalent. Moreover, the value of 
$\chi_1$ will agree with (\ref{chiq}). Having made the gluon Schwinger Dyson
equation finite, it is elementary to expand (\ref{gSDfin}) to $O(1/\Nf^2)$ to 
obtain an expression for $z_2$. We find 
\begin{eqnarray} 
\frac{z_2}{z_1} &=& (\eta_1 + \chi_1) \left[ \Psi(\mu) + \frac{1}{2(\mu-1)}
+ \frac{1}{(\mu-2)} + \frac{2}{(2\mu-3)} \right] \nonumber \\ 
&& -~ \frac{(2\mu-1)}{8(\mu-1)} \left[ \frac{4\eta_1}{(2\mu-1)^2} ~+~ 
C_A \frac{(2\mu-3)\Gamma(\mu)\tilde{y}_1}{4(2\mu-1)z_1} \right. 
\nonumber \\
&& \left. ~~~~~~~~~~~~~~~~+~ \left(C_F - \frac{C_A}{2}\right) z_1 \left( 
\Pi^\prime_1 + \frac{\Xi^\prime_1}{2(2\mu-1)} 
+ \frac{X_1}{2(2\mu-1)^2} \right) \right. \nonumber \\ 
&& \left. ~~~~~~~~~~~~~~~~-~ C_A z^2_1 \left( \Pi^\prime_2 + 
\frac{\Xi^\prime_2}{2(2\mu-1)} + \frac{X_2}{(2\mu-1)^2} \right) \right] ~.  
\end{eqnarray}  
Therefore with this value for $z_2$ we can establish a formal expression for
$\eta_2$ in arbitrary gauge which depends only on the values of the various
graphs of figures \ref{qSDfig}, \ref{gSDfig} and \ref{ghSDfig} by eliminating 
$z_2$ in (\ref{eta2defn}). For completeness, we quote the coordinate space 
values of the integrals which will determine $\eta_2$ from these equations. 
First, we recall the graphs of the QED sector are 
\begin{eqnarray}  
\Sigma_1 &=& \frac{4}{\mu(2\mu-3)^2\Gamma^2(\mu)} 
\left[ \frac{2(2\mu-1)^2(\mu-2)^2}{\Delta} ~+~ 4(\mu-1)^2 \left( 
\frac{2(\mu-1)^2}{\mu} + \frac{(2\mu-3)}{2} \right) \right. \nonumber \\ 
&& \left. ~~~~~~~~~~~~~~~~~~~~~~~~-~ 4(2\mu-1) ~-~ (2\mu-1)(\mu-2) ~-~ 
\frac{8(2\mu-1)^2(\mu-2)^2}{(2\mu-3)} \right] \\ 
\Pi_1 &=& -~ \frac{\Xi_1}{2} ~=~ -~ \frac{16}{(2\mu-3)\Gamma^2(\mu)} 
\left[ \frac{(2\mu-1)(\mu-2)}{\mu\Delta} \right. \nonumber \\ 
&& \left. ~~~~~~~~~~~~~+~ \frac{3}{(\mu-1)} ~-~ 3(\mu-1)\Theta(\mu) ~-~ 
\frac{2(2\mu-1)(\mu-2)}{\mu(2\mu-3)} \right] ~.  
\end{eqnarray}  
For the three loop graphs we have 
\begin{eqnarray} 
\Sigma_2 &=& -~ \frac{8(2\mu-1)^2(\mu-2)}{(2\mu-3)^3\mu^2 \Gamma^4(\mu) \eta_0}
\left[ \frac{(2\mu-1)(\mu-2)(\mu-1)}{\Delta} ~-~  6\mu(\mu-1)(\mu-2)\Theta(\mu)
\right. \nonumber \\ 
&& \left. ~~~+~ [8\mu^6 - 48\mu^5 + 118\mu^4 - 130\mu^3 + 34\mu^2 
+ 35\mu - 12]/[\mu(\mu-1)(2\mu-3)] \frac{}{} \! \right] 
\end{eqnarray} 
and 
\begin{eqnarray} 
\Pi_2 &=& -~ \frac{8}{(2\mu-3)^2\mu^2\Gamma^4(\mu)\eta_0} \left[ 
\frac{2(2\mu-1)^2(\mu-1)(\mu-2)}{\Delta} \right. \nonumber \\ 
&& \left. ~~~~~+~ 4\mu(2\mu-1)(\mu-1)(\mu-2)[\Phi(\mu) + \Psi^2(\mu) 
- 6\Theta(\mu)] \right. \nonumber \\
&& \left. ~~~~~-~ \frac{2\mu\Psi(\mu)}{(2\mu-3)} 
(8\mu^4-76\mu^3+198\mu^2-193\mu+62) \right. \nonumber \\ 
&& \left. ~~~~~+~ 8(4\mu^7-34\mu^6+123\mu^5-222\mu^4+197\mu^3 
-78\mu^2+16\mu-3) \right. \nonumber \\ 
&& \left. ~~~~~~~~~~/[\mu(\mu-1)(2\mu-3)] \frac{}{} \right] ~.  
\end{eqnarray} 
The longitudinal component is related to $\Pi_2$ by 
\begin{equation} 
\Xi_2 ~=~ -~ 2\Pi_2 ~+~
\frac{64}{(2\mu-3)^2\mu^2\Gamma^4(\mu)\eta_0} 
\left[ (2\mu-1)(\mu-2) ~+~ \frac{\mu(\mu-1)^2a^3(1)} 
{2(2\mu-3)a(3-\mu)} \right] ~.  
\end{equation} 
Therefore, substituting the values of the integrals in the formal expressions
for $\chi_1$ and $m_1$, we have explicitly 
\begin{eqnarray} 
m_1 &=& -~ \frac{C_F\eta_0}{2T_F} ~-~ \frac{(2\mu-1)(\mu-3) C_A \eta_0} 
{8(2\mu-1)(\mu-2)T_F} \\  
\chi_1 &=& -~ \frac{C_F\eta_0}{T_F} ~-~ \frac{C_A\eta_0}{2(\mu-2)T_F} ~.  
\end{eqnarray} 

The relation between the coordinate space variables and the momentum space ones
of earlier sections is determined from the Fourier transform of the asymptotic
scaling forms. For instance, it is easy to deduce that the respective quark 
amplitudes $A$ and $\tilde{A}$ are related by 
\begin{equation} 
\tilde{A} ~=~ -~ \frac{ia(\alpha-1)A}{(\alpha-1)} 
\end{equation} 
whilst the respective gluon amplitudes are determined from comparing the 
coefficients of the $\eta_{\mu\nu}$ component. We find 
\begin{equation} 
\tilde{B} ~=~ \frac{2(\mu-\beta)a(\beta)B}{(2\mu-2\beta-1)} ~. 
\end{equation} 
Thus, 
\begin{equation} 
\tilde{z} ~=~ -~ \frac{(\mu-\beta) a^2(\alpha-1)a(\beta)z}{(\alpha-1)^2 
(2\mu-2\beta-1)} 
\end{equation} 
which implies 
\begin{equation} 
\tilde{z}_1 ~=~ -~ \frac{2z_1}{(2\mu-3)\Gamma(\mu)} 
\end{equation} 
and 
\begin{equation} 
\tilde{z}_2 ~=~ -~ \frac{2}{(2\mu-3)\Gamma(\mu)} \left[ z_2 ~-~ 
\frac{2z_1\eta_1}{(2\mu-3)} \right] 
\end{equation} 
upon expanding in powers of $1/\Nf$. Moreover, variables involving the ghost 
amplitudes are related by $\tilde{y}$ $=$ $a^2(\gamma)a(\beta)y$. Similarly, we
can relate the values of the three loop integrals expressed in either 
coordinate or momentum space. For instance, if the value of the three loop 
gluonic graph in coordinate space is 
\begin{equation} 
\frac{1}{(x^2)^{2\mu-1-2\Delta}} \left[ \Pi_2 \eta_{\mu\nu} ~+~ \Xi_2 
\frac{x_\mu x_\nu}{x^2} \right] 
\end{equation} 
and in momentum space  
\begin{equation} 
\frac{1}{(p^2)^{1-\mu+2\Delta}} \left[ \tilde{\Pi}_2 \eta_{\mu\nu} ~+~ 
\tilde{\Xi}_2 \frac{p_\mu p_\nu}{p^2} \right] 
\end{equation} 
then we have, 
\begin{eqnarray} 
\Pi_2 &=& \frac{4(\mu-1+\Delta)^2 a^6(\mu-1) a^2(1-\Delta)} 
{(\mu-1)^6 (2\mu-3+2\Delta)^2 a(2\mu-1-2\Delta)} \left[
\tilde{\Pi}_2 ~-~ \frac{\tilde{\Xi}_2}{2(\mu-1-2\Delta)} \right] \nonumber \\
\Xi_2 &=& \frac{4(\mu-1+\Delta)^2(2\mu-1-2\Delta) a^6(\mu-1) a^2(1-\Delta) 
\tilde{\Xi}_2}{(\mu-1-2\Delta)(\mu-1)^6(2\mu-3+2\Delta)^2 a(2\mu-1-2\Delta)} ~. 
\end{eqnarray} 
Likewise, 
\begin{equation} 
\Sigma_2 ~=~ \frac{(\mu-3\Delta)(\mu-1+\Delta)^3 a^5(\mu-1)
\tilde{\Sigma}_2}
{16(\mu-1)^5(2\mu-3+2\Delta)^3 a^3(\mu-1+\Delta) a(\mu-3\Delta)} ~.  
\end{equation}  
So, for example, the finite parts of the momentum space integrals are given by 
\begin{eqnarray} 
\tilde{\Pi}^\prime_2 &=& \frac{8(\mu-1)^3 a^4(1)a^2(2\mu-2)} 
{(2\mu-1)^2\Gamma(\mu)} \left[ 4 \left( \Psi^2(\mu) + \Phi(\mu) - 6\Theta(\mu) 
\right) \right. \nonumber \\
&& \left. -~ \frac{a^3(1)}{(2\mu-1)(2\mu-3)(\mu-2)a(3-\mu)} \right. 
\nonumber \\ 
&& \left. +~ \frac{(32\mu^7-176\mu^6+524\mu^5-952\mu^4+927\mu^3-468\mu^2 
+180\mu-48)}{(\mu-1)^2(2\mu-1)(2\mu-3)(\mu-2)\mu^2} \right. \nonumber \\  
&& \left. -~ \frac{4}{(\mu-1)} \left( \Psi(\mu)  
+ \frac{(4\mu^5-16\mu^4+32\mu^3-40\mu^2+25\mu-6)} 
{(2\mu-1)(2\mu-3)(\mu-1)(\mu-2)\mu^2} \right) ~-~ \frac{1}{(\mu-1)^2} 
\right] \\  
\tilde{\Xi}^\prime_2 &=& -~ \tilde{\Pi}^\prime_2 ~-~ \frac{8a^4(1)a^2(2\mu-2)} 
{(2\mu-1)^2\Gamma(\mu)} \left[ \frac{}{} 2(\mu-1) \right. \nonumber \\ 
&& \left. -~ \frac{(\mu-1)^3}{(2\mu-3)(\mu-2)\mu} \left( 4(2\mu-3)(\mu-2) 
- \frac{\mu a^3(1)}{a(3-\mu)} \right) \right]  
\end{eqnarray}  
and 
\begin{equation} 
\tilde{\Sigma}^\prime_2 ~=~ \frac{2(\mu-1)(2\mu-1)^2(\mu-2)^2} 
{\mu^2\Gamma^3(\mu)\eta_0} \left[ \Theta(\mu) ~-~ 
\frac{(4\mu^5-6\mu^4-13\mu^3+40\mu^2-40\mu+10)}{6\mu^2(\mu-1)^2(\mu-2)} 
\right] ~. 
\end{equation} 

\sect{Mass anomalous dimension.}
Having presented the formalism to calculate the quark anomalous dimension at 
$O(1/\Nf^2)$ we are now turn to the quark mass anomalous dimension at the same 
order. Perturbatively the mass anomalous dimension is known to four loop 
accuracy. Thus comparison of results obtained in both perturbation theory and 
the $1/\Nf$ expansion will provide a non-trivial test on the validity of our 
results since, for instance, only three loop diagrams are considered at 
$O(1/\Nf^2)$ in the large $\Nf$ formalism. To complete the calculation we 
discuss some observations here which allow us to relate the values of most of 
the graphs contributing to the mass dimensions at $O(1/\Nf^2)$ to those 
used for the quark anomalous dimensions. It transpires that this greatly
reduces the amount of calculation and will provide a method for tackling the 
evaluation of the anomalous dimensions of other operators. 

As is well known the mass anomalous dimension coincides with the anomalous 
dimension of the mass operator $\bar\psi\psi$,
\begin{equation}
\gamma_m ~=~ \gamma_{\bar\psi\psi} ~+~ 2\gamma_\psi ~.
\end{equation}
Here $\gamma_\psi$ is the quark anomalous dimension and the RG function 
$\gamma_{\bar\psi\psi}$ is expressed via the renormalization constant of the 
mass operator by 
\begin{equation}
\gamma_{\bar\psi\psi} ~=~ M\partial_M \ln Z_{\bar\psi\psi} ~, \ \ \ \ 
[\bar\psi\psi]_R ~=~ Z_{\bar\psi\psi} Z^{-2}_{\psi} \, [\bar\psi_0\psi_0]_0 ~.
\end{equation}
To determine the renormalization constant
$Z_{\bar\psi\psi}$ at second order in
the $1/\Nf$ expansion one has to compute the divergences of the one-particle
irreducible 2-point Green function with the insertion of the mass operator. The
corresponding diagrams can be obtained from the diagrams contributing to the 
quark self-energy of figure \ref{fig1} by the insertion of a $\bar\psi\psi$ 
vertex in the quark lines connecting the external vertices. At the level of 
diagrams this equivalent to the insertion of the unit operator in a fermion 
line 
\[
\frac{\pslash}{p^2} ~\rightarrow~  
\frac{\pslash}{p^2}\cdot\frac{\pslash}{p^2} ~. 
\]
As was discussed in section 4 the differentiation of the propagator diagrams 
with respect to $p_\mu$, where $p$ is the external momentum, results in the 
same set of the diagrams as for the mass operator. The only difference is that 
the inserted vertex in this case is $(-\gamma_\mu)$. We shall exploit this fact
and show below that provided the Landau gauge is used then the corresponding 
contributions to $Z_1$ $=$ $Z_\psi^2$ and $Z_{\bar\psi\psi}$ from all diagrams 
from both sets, except for two which arise from graphs~$(b)$ and $(c)$ of  
figure \ref{fig1} when the insertion of the new vertex is in the middle quark 
line, are related to one other by
\begin{equation}
\label{relation}
Z_{\bar\psi\psi}^{(i)} ~=~ \frac{\mu Z_{1}^{(i)}}{(\mu - 2)} ~.
\end{equation}  
So given that $\gamma_\psi$ is known then to determine the mass dimension 
$\gamma_m$ at $O(1/\Nf^2)$ it is sufficient to compute the contributions to 
$Z_1$ and $Z_{\bar\psi\psi}$ from the two extra diagrams. However, their
calculation does not lead to any difficulties using the methods discussed in 
section 4.  

\begin{figure}[t]
\centerline{\epsfxsize12.0cm\epsfbox[200 230  500 300]{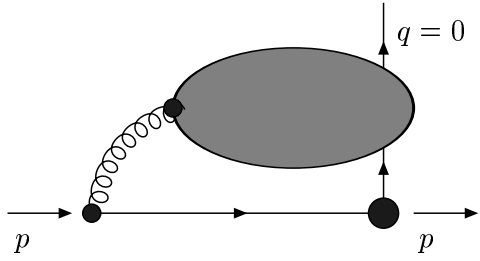}}
\vspace{1.5cm} 
\caption[]{External momenta routing in the quark $2$-point function with an  
operator insertion. The large black dot stands for the $V_i$ $=$ 
$\{I,(-\gamma_\mu)\}$.}
\label{fig3}
\end{figure}

We now proceed to the proof of the relation~(\ref{relation}). First, we use the
freedom we have to choose the external momentum flow in a diagram arbitrarily
when calculating the contribution to the renormalization constants by directing
the external momentum flow through the inserted vertex and out through the 
nearest external vertex. After this all the diagrams in question take the form
shown in figure \ref{fig3}. 
All that one needs to know about the coloured block in figure \ref{fig3} is its 
Lorentz structure,
\begin{equation}
\label{gvertex}
\Gamma_\nu(q,\Delta) ~=~ \frac{1}{(q^2)^{n_i\Delta}} \left(
\gamma_\nu A(\Delta)+\frac{q_\nu\qslash}{q^2} B(\Delta)\right ) ~,
\end{equation}
where $n_i$ is the number of the gluon lines in the block and the exact form of
the functions $A(\Delta) and B(\Delta)$ are irrelevant for our present 
discussion. Since the gluon propagator which joins this generalized vertex is 
purely transverse in the Landau gauge the second term in (\ref{gvertex}) will 
not contribute to the final answer. Thus the momentum integral for the diagram 
in figure \ref{fig3} can be cast in the form
\begin{equation}
A(\Delta)\int \frac{{\rm d}^{2\mu}q}{(2\pi)^{2\mu}}
\frac{\gamma^{\rho}\,(\pslash - \qslash)\,V_i\,\gamma^{\nu}
\,\qslash\,P^{\perp}_{\rho\nu}(q)} {(p-q)^2\, (q^2)^{\mu+(n_i+1)\Delta}} ~. 
\label{intf3}
\end{equation}
We are interested in the pole terms of this integral only, and bearing in mind 
that after the subtraction of the divergent subgraphs, which also have the same
form, the latter is independent of $p$. Therefore one can rewrite~(\ref{intf3})
as 
\begin{equation}
-~A(\Delta) \left (V_i- \frac{1}{2\mu} \gamma_\rho\gamma_\nu V_i 
\gamma^\nu\gamma^\rho\right) \int \frac{{\rm d}^{2\mu}q}{(2\pi)^{2\mu}}
\frac{1}{(p-q)^2\, (q^2)^{\mu-1-(n_i+1)\Delta}} ~, 
\label{intF}
\end{equation}
where we have used the fact that up to $p$-dependent terms one can replace
$q_\rho\,q_\nu$ by $q^2/(2\mu)$ in the integrand. It is easy to deduce that the
combination in the prefactor in~(\ref{intF}) is equal to $-(2\mu-1)$ for 
$V$~$=$~$I$, and to $(2\mu-1)(\mu-2)/\mu$ for $V$~$=$~$-$~$\gamma_\mu$. Then 
one can immediately conclude that irrespective of the exact form of the 
constant $A(\Delta)$ which corresponds to the structure of the coloured block 
in figure \ref{fig3}, the contributions of the corresponding diagram to the 
renormalization constants $Z_1$ and $Z_{\bar\psi\psi}$ are indeed related by 
equation~(\ref{relation}). To complete the calculation of $\gamma_m$ one needs 
to compute the contributions from the graphs originating from those $b$, $c$ in
figure \ref{fig1}. These, as well as the determination of the graphs when the 
gluon propagator is taken in an arbitrary gauge, can readily be performed. 

\sect{Discussion.} 

We are now in a position to assemble our results. First, using the formalism of
section 3 we record that the $d$-dimensional expression for the quark anomalous 
dimension is 
\begin{eqnarray}
\gamma_\psi^{(2)} &=& \frac{C_F\eta_0^2}{2(\mu-2)(2\mu-1)} \left[ C_F \left( 
\frac{2(\mu-1)^2(\mu-3)}{\mu(\mu-2)} ~+~ 3\mu(\mu-1) \left(
\Theta(\mu)-\frac{1}{(\mu-1)^2} \right) \right) \right. \nonumber \\
&&\nonumber\\
&&+~ \frac{C_A}{2} \left( \frac{(12\mu^4-72\mu^3+126\mu^2-75\mu+11)}{(2\mu-1) 
(2\mu-3)(\mu-2)} ~-~ \mu(\mu-1)\left(\Psi^2(\mu)+\Phi(\mu)\right) \right. 
\nonumber \\
&& +~ \frac{(8\mu^5-92\mu^4+270\mu^3-301\mu^2 +124\mu-12)\Psi(\mu)} 
{2(\mu-2)(2\mu-1)(2\mu-3)} \nonumber \\
&&\nonumber \\
&&\left. \left. -~ \frac{\xi(\mu^2-4\mu+2)}{2(\mu-1)(\mu-2)} ~-~
\frac{\xi^2(2\mu^3-11\mu^2+12\mu-4)}{4(\mu-1)(\mu-2)(2\mu-1)} \right) 
\right] ~.
\label{eta2gen} 
\end{eqnarray} 
We note that as an initial check on the result setting $C_F$~$=$~$1$ and 
$C_A$~$=$~$0$ one obtains the expression for the quark anomalous dimension of 
the abelian Thirring model. Again, in agreement with general results,
\cite{Zinn}, we find that in this case $\gamma_\psi^{(2)}$ is independent of 
the gauge fixing parameter. Further, collecting all the results for the quark 
mass anomalous dimension, we find  
\begin{eqnarray} 
\gamma_{m,2} &=& -~ 2\left[ \frac{2(\mu-1)^2(\mu-3)}{\mu(\mu-2)} ~+~ 3\mu 
(\mu-1) \left( \Theta(\mu) - \frac{1}{(\mu-1)^2} \right) \right. \nonumber \\ 
&& \left. ~~~~~~+~ \frac{(2\mu^2-4\mu+1)}{(\mu-2)} \right] 
\frac{C^2_F \eta_0^2}{(\mu-2)^2(2\mu-1)} \nonumber \\
&& -~ \left[ \frac{(12\mu^4-72\mu^3+126\mu^2-75\mu+11)} 
{(2\mu-1)(2\mu-3)(\mu-2)} ~-~ \mu(\mu-1) \left( \Psi^2(\mu)+\Phi(\mu) \right) 
\right. \nonumber \\ 
&& \left. ~~~~~+~ \frac{(8\mu^5-92\mu^4+270\mu^3-301\mu^2+124\mu-12)\Psi(\mu)} 
{2(2\mu-1)(2\mu-3)(\mu-2)} \right. \nonumber \\ 
&& \left. ~~~~~-~ \frac{\mu^2(2\mu-3)^2}{4(\mu-2)(\mu-1)} \right] 
\frac{C_F C_A \eta_0^2}{(\mu-2)^2(2\mu-1)} ~, 
\label{gam_m_2} 
\end{eqnarray}
where $\gamma_m(g_{*})$ $=$ $\sum_{r=1}^\infty \gamma_{m,r}/n^r$. We have
performed the calculations with a non-zero gauge parameter to check that it 
cancels exactly since the mass dimension is a gauge independent quantity. Aside
from the internal checks on the integration which we discussed earlier, there 
are other checks on the correctness of (\ref{eta2gen}) and (\ref{gam_m_2}). 
First, the terms of (\ref{eta2gen}) and (\ref{gam_m_2}) involving $C^2_F$ have 
both been derived previously in QED in \cite{JG94} and \cite{JG93m} 
respectively. However, a more stringent check rests in the relation of the 
results with the critical renormalization group equation since the RG functions
evaluated at $g_*$ ought to agree with (\ref{eta2gen}) and (\ref{gam_m_2}) at 
the same order in $1/\Nf$ and $O(\epsilon^4)$. It is elementary to verify that 
our results are in agreement with the Landau gauge anomalous dimension of 
\cite{LV93,CR99}. Also the result for $\gamma_m(g_*)$ is consistent with 
\cite{NRT79,T82,VLR97,C97}. 

Having verified this agreement we can now determine new information on the 
structure of the renormalization group functions in principle to all orders in 
the strong coupling constant. However, we will record values of the
coefficients to six loops by writing 
\begin{eqnarray} 
\gamma(a) &=& -~ \left( \frac{1}{4} T_F \Nf + \frac{3}{16} C_F - \frac{25}{32} 
C_A \right) C_F a^2 \nonumber \\
&& +~ \left( 320 T^2_F \Nf^2 + 432 C_F T_F \Nf - 4592 C_A T_F \Nf + 216 C^2_F  
\right. \nonumber \\ 
&& \left. ~~~~+~ (1728\zeta(3) - 5148) C_F C_A + (9155 - 1242\zeta(3)) 
C^2_A \right) \frac{C_F a^3}{4608} \nonumber \\  
&& +~ \sum_{r=4}^\infty \left( c_{r0} T_F^{r-1} \Nf^{r-1} + c_{r1} T_F^{r-2}  
\Nf^{r-2} \right) C_F a^r ~+~ O \left( \frac{1}{\Nf^3} \right) 
\label{anomdimgen} 
\end{eqnarray} 
for the Landau gauge and
\begin{eqnarray} 
\gamma_m(a) &=& -~ \frac{3}{2} C_F a ~+~ \left( \frac{10}{24} T_F \Nf 
- \frac{3}{16} C_F - \frac{97}{48} C_A \right) C_F a^2 \nonumber \\
&& +~ \sum_{r=3}^\infty \left( m_{r0} T_F^{r-1} \Nf^{r-1} + m_{r1} T_F^{r-2}  
\Nf^{r-2} \right) C_F a^r ~+~ O \left( \frac{1}{\Nf^3} \right) 
\label{massdimgen} 
\end{eqnarray} 
for the mass dimension where the order symbol means that contributions which
are third order in $1/\Nf$ and $O(\epsilon^7)$ at criticality are ignored. 
Although the full four loop $\MSbar$ $\beta$-function is only available it may 
seem that it is not possible to decode the Landau gauge information in 
(\ref{eta2gen}) and (\ref{gam_m_2}) beyond four loops since this requires $a_*$
to $O(\epsilon^6)$. However, recalling that we are computing within the $1/\Nf$
expansion we can use the information contained in the QCD $\beta$-function 
exponent $\omega$~$=$~$-$~$\beta^\prime(a_*)/2$ which has been determined at 
$O(1/\Nf)$ in \cite{JG96}. From \cite{JG96}, we deduce that the relevant 
corrections are  
\begin{eqnarray}
a_{*} &=& \frac{3\epsilon}{T_F \Nf} ~+~ \left[ \frac{33}{4} C_A\epsilon ~-~ 
\left( \frac{27}{4}C_F+\frac{45}{4}C_A \right) \epsilon^2 ~+~ 
\left( \frac{99}{16}C_F + \frac{237}{32}C_A \right) \epsilon^3 \right. 
\nonumber \\ 
&& \left. ~~~~+~ \left( \frac{77}{16}C_F + \frac{53}{32}C_A \right) 
\epsilon^4 ~-~ \frac{3}{256} \left[ (288\zeta(3) + 214)C_F 
+ (480\zeta(3)-229)C_A \right] \epsilon^5 \right. \nonumber \\ 
&& \left. ~~~~-~ \frac{1}{256} \left[ (3168\zeta(3) - 2592\zeta(4) 
- 1506) C_F \right. \right. \nonumber \\ 
&& \left. \left. ~~~~~~~~~~~~~~+~ (3792\zeta(3) - 4320\zeta(4) + 1359) 
C_A \right] \epsilon^6 ~+~ O(\epsilon^7) \right] \frac{1}{T_F^2 \Nf^2} ~+~ 
O \left( \frac{1}{\Nf^3} \right) ~. \nonumber \\  
\label{acritfull}
\end{eqnarray}
Therefore, with this value we deduce the new exact contributions to 
(\ref{anomdimgen}) and (\ref{massdimgen}) are 
\begin{eqnarray} 
c_{40} &=& \frac{35}{1296} \nonumber \\ 
c_{41} &=& \frac{C_F}{72}[19 - 18\zeta(3)] ~+~ \frac{C_A}{1152} [293 
+ 288\zeta(3)] \nonumber \\ 
c_{50} &=& \frac{1}{7776} [83 - 144\zeta(3)] \nonumber \\ 
c_{51} &=& \frac{C_F}{31104}[5616\zeta(3) - 3888\zeta(4) - 659] ~+~ 
\frac{C_A}{62208}[7776\zeta(4) - 8928\zeta(3) - 1783] \nonumber \\  
c_{60} &=& \frac{1}{15552} [65 - 144\zeta(4) + 80\zeta(3)] \nonumber \\ 
c_{61} &=& \frac{C_F}{46656}[4212\zeta(4) - 2592\zeta(5) - 432\zeta(3) - 2219]
\nonumber \\ 
&& +~ \frac{C_A}{31104}[4608\zeta(5) - 1440\zeta(4) -4744\zeta(3) - 925] 
\end{eqnarray} 
and 
\begin{eqnarray} 
m_{50} &=& \frac{5\zeta(3)}{162} - \frac{\zeta(4)}{18} + \frac{65}{2592} \\ 
m_{51} &=& \left[ \frac{5\zeta(4)}{8} - \frac{\zeta(5)}{3} ~-~ 
\frac{11\zeta(3)}{48} - \frac{4483}{20736} \right] C_F \nonumber \\
&& +~ \left[ \frac{8\zeta(5)}{9} - \frac{17\zeta(4)}{36} 
- \frac{671\zeta(3)}{1296} - \frac{18667}{124416} \right] C_A \nonumber \\ 
m_{60} &=& \frac{1}{46656}[560\zeta(3) + 720\zeta(4) - 1728\zeta(5) + 451] 
\nonumber \\ 
m_{61} &=& \frac{C_F}{186624}[46704\zeta(3) - 23328\zeta(4) + 51840\zeta(5)
- 41472\zeta^2(3) - 25920\zeta(6) - 12283] \nonumber \\ 
&& +~ \frac{C_A}{186624}[56728\zeta(3) - 59472\zeta(4) - 112896\zeta(5)
+ 55296\zeta^2(3) + 112320\zeta(6) - 22709] ~. \nonumber \\ 
\end{eqnarray} 
We have given the explicit expression for $c_{41}$ in order to compare with the
full $\MSbar$ gauge calculation of \cite{CR99} which was carried out for the 
particular colour group $SU(\Nc)$ and note exact agreement. Further, we note 
that in \cite{CDGM} it was pointed out how the asymptotic Pad\'{e} approximant 
predictions of \cite{EJJKS98} for $m_{50}$ and $m_{51}$ compared with the exact
values we have determined. 

Although we have concentrated on the relation of our results to four 
dimensional perturbation theory, we can also quote values of the exponents in
three dimensions. These will be important for alternative critical point 
investigations of the non-abelian Thirring model and QCD equivalence, such as
the lattice. For the wave function, we find from (\ref{eta2gen}) 
\begin{equation} 
\eta ~=~ \frac{4C_F(3\xi - 2)}{3\pi^2 T_F \Nf} ~+~ \frac{8C_F} 
{9\pi^4 T^2_F \Nf^2} \left[ \left( 64 - 6\pi^2 \right) C_F 
+ \left( 8\xi^2 + 14\xi - 18 - \pi^2 \right) C_A \right] ~+~ 
O \left( \frac{1}{\Nf^3} \right) ~. 
\end{equation} 
However, for the physically more interesting mass exponent we deduce,  
\begin{equation} 
\gamma_m ~=~ -~ \frac{32C_F}{3\pi^2 T_F \Nf} ~+~ \frac{32C_F [(56-6\pi^2)C_F 
- (18 + \pi^2)C_A]}{9\pi^4 T^2_F \Nf^2} ~+~ O \left( \frac{1}{\Nf^3} \right) ~. 
\end{equation} 
It would be interesting to see how the numerical values of these exponents 
compare with the lattice. Indeed the dominant contribution from the 
$O(1/\Nf^2)$ correction in the latter exponent arises from the $C_A$ term. 

In conclusion we have provided an elegant alternative to computing perturbative
information on QCD at a new order in the large $\Nf$ expansion. Although the
evaluation of the Feynman integrals has formed a significant part of the 
discussion, is technically quite involved and provides useful techniques for
future massless integral evaluation, we believe we have demonstrated that the 
computation of other renormalization group functions in QCD is viable in 
principle. An example of this is the twist-$2$ operators which arise in deep
inelastic scattering. Furthermore, the relation of the NATM to QCD has been 
put on a firmer ground. Indeed we believe that this critical relation between
both models deserves further investigation. 

\vspace{1cm}  
\noindent 
{\bf Acknowledgements.} This work was carried out with the support of the
Russian Foundation of Basic Research, Grant 97--01--01152 (SED and ANM), 
by DFG Project N KI-623/1-2 (SED), a University of Pisa Exchange Research 
Fellowship (MC), PPARC through an Advanced Fellowship (JAG) and GSI (ANM). Also
SED and ANM would like to thank Prof.~A.N.~Vasil'ev for useful discussions. 
Invaluable to the calculations were the symbolic manipulation programme {\sc 
Form}, \cite{form}, and computer algebra package {\sc Reduce}, \cite{reduce}. 

\vspace{0.8cm} 
\appendix
\sect{Basic integration rules.} 
In this appendix we give formul{\ae} for the Fourier transformation, 
integration of chains and uniqueness relations for the vertices of different 
type. The following notation is used. We define 
\[
a(x_1,\ldots,x_n) ~=~ \prod_{i=1}^n a(x_i) 
\] 
where $a(\alpha)$ $=$ $\Gamma(\alpha^\prime)/\Gamma(\alpha)$ with 
$\alpha^\prime$ $=$ $(\mu$ $-$ $\alpha)$. The Fourier transforms of various 
functions are given by  
\begin{eqnarray}
&&\int d^{{2\mu}}x\> \frac{e^{i px}}{(x^2)^\al} ~=~ 
\pi^\mu a(\al) \frac{2^{2\al^\prime}}{(p^2)^{\al^\prime}}\nonumber\\
&&\nonumber\\
&&\int d^{{2\mu}}x\> e^{i px} \frac{x_{\nu}}{(x^2)^{\al+1}} ~=~ 
i\pi^\mu 2^{2\al^\prime-1}\frac{a(\al)}{\al}
\frac{p_{\nu}}{(p^2)^{\al^\prime}}\nonumber\\
&&\nonumber\\
&&\int d^{{2\mu}}x\> e^{i px} \frac{x_{\mu}x_{\nu}}{(x^2)^{\al+1}} ~=~ 
\pi^\mu\frac{a(\al)}{\al} \frac{2^{2\al^\prime-1}}{(p^2)^{\al^\prime}}
\left[\eta_{\mu\nu}-2\al^{\prime}\frac{p_{\mu}p_{\nu}}{p^{2}} \right] \\
&&\nonumber\\
&&\int d^{{2\mu}}x\> e^{i px} \frac{x_{\mu}x_{\nu}x_{\rho}}{(x^2)^{\al+1}} ~=~ 
i\pi^\mu{a(\al)} \frac{\al^\prime}{\al} \frac{2^{2\al^\prime}} 
{(p^2)^{\al^\prime+1}} \left[ 
p_{\mu}\eta_{\nu\rho}+p_{\nu}\eta_{\mu\rho}+p_{\rho}\eta_{\mu\nu} 
- 2(\al^\prime+1)\frac{p_{\mu}p_{\nu}p_{\rho}}{p^2} \right] \nonumber \\
&&\nonumber\\&&\int d^{{2\mu}}x\> \frac{e^{i px}}{(x^2)^\al}
\left[ {\eta_{\mu\nu}-\frac{2x_{\mu}x_{\nu}}{x^{2}}} \right] ~=~ 
\pi^\mu\frac{a(\al)}{\al} \frac{2^{2\al^\prime}}{(p^2)^{\al^\prime}}
\left[ (\al-1)\eta_{\mu\nu}+2\al^{\prime}\frac{p_{\mu}p_{\nu}}{p^{2}} \right] 
\nonumber \\
&& \nonumber \\
&&\int \frac{d^{{2\mu}}k}{(2\pi)^{2\mu}}\>{e^{i px}}
\frac{P_{\mu\nu}^{\perp}}{(k^2)^\al} ~=~ 
\frac{(2\al-1)}{(4\pi)^\mu}\frac{a(\al)}{\al}
\frac{2^{2\al^\prime-1}}{(x^2)^{\al^\prime}}
\left (\eta_{\mu\nu}+\frac{2\al^\prime}{(2\al-1)}\frac{x_{\mu}x_{\nu}}{x^{2}} 
\right) ~. 
\end{eqnarray}
The integration of the basic chain with various tensor generalizations are 
given by  
\begin{eqnarray}
&&\int \frac{{\rm d}^{2\mu}x}{\pi^{\mu}}
\frac{1}{((z-x)^2)^\alpha\,(x^2)^\beta} ~=~ 
\frac{a(\alpha,\beta,\alpha^{\prime}+\beta^{\prime})}
{z^{2(\alpha+\beta-\mu)}} ~, \nonumber \\
&&\nonumber \\
&&\int \frac{{\rm d}^{2\mu}x}{\pi^{\mu}}
\frac{x_\nu}{((z-x)^2)^\alpha\,(x^2)^\beta} ~=~ 
\frac{a(\alpha,\beta,\alpha^{\prime}+\beta^{\prime})\beta^{\prime}}
{\alpha^{\prime}+\beta^{\prime}}
\frac{z_{\nu}}{(z^2)^{\alpha+\beta-\mu}} ~, \nonumber \\
&&\nonumber \\
&&\int \frac{{\rm d}^{2\mu}x}{\pi^{\mu}}
\frac{x_{\nu}(z-x)_{\mu}}{((z-x)^2)^\alpha\,(x^2)^\beta} ~=~ 
-\, \frac{\alpha^\prime\beta^\prime \,  
a(\alpha,\beta,\alpha^\prime+\beta^\prime+1)} 
{2(\alpha^\prime+\beta^\prime+1) \, (z^2)^{\alpha+\beta-\mu-1}}
\left[\delta_{\mu\nu}+ 2(\alpha^{\prime}+1-\beta)\frac{z_{\mu}z_{\nu}}{z^2} 
\right] \, , \nonumber \\
&&\nonumber \\
&& \int \frac{{\rm d}^{2\mu}x}{\pi^{\mu}}
\frac{\xslash(\zslash-\xslash)}{((z-x)^2)^\alpha\,(x^2)^\beta} ~=~ -~  
\alpha^\prime\beta^\prime\,a(\alpha,\beta,\alpha^\prime+\beta^\prime+1)
\frac{1}{(z^2)^{\alpha+\beta-\mu-1}} ~, \\
&&\nonumber \\
&&\int \frac{{\rm d}^{2\mu}x}{\pi^{\mu}}
\frac{1}{((z-x)^2)^\alpha \, (x^2)^\beta}
\left[A\delta_{\mu\nu} ~+~ B \frac{x_{\mu}x_{\nu}}{x^2}\right] \nonumber \\
&&\ \ \ \ \ \ \ ~=~  
\frac{a(\alpha,\beta,\alpha^{\prime}+\beta^{\prime})}{(z^2)^{\alpha+\beta-\mu}}
\left[\Biggl(A+\frac{B\alpha^{\prime}}{2\beta(\alpha^{\prime}+\beta^{\prime})} 
\Biggr) \delta_{\mu\nu} ~+~ \frac{B\beta^{\prime}(\alpha-\beta^{\prime})}
{\beta(\alpha^{\prime}+\beta^{\prime})} \frac{z_{\mu}z_{\nu}}{z^2}\right] ~. 
\end{eqnarray}
Next we list the uniqueness relations for the three type of vertices which 
arose in our calculations where the appropriate uniqueness value for each 
vertex is recorded in parentheses beside each rule.  

\noindent
Boson vertex ($\alpha$ $+$ $\beta$ $+$ $\gamma$ $=$ $2\mu$).

\centerline{\epsfxsize12.0cm\epsfbox[140 150  450 300]{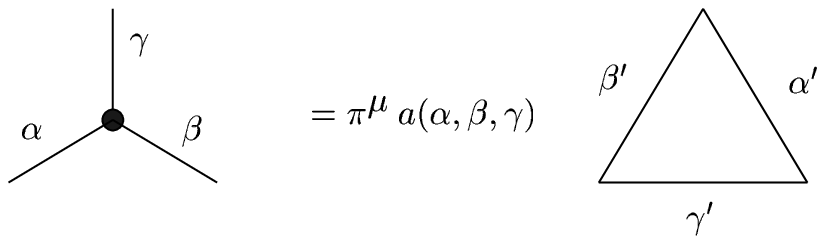}}
\vskip -1cm

\noindent
Gross-Neveu vertex ($\alpha$ $+$ $\beta$ $+$ $\gamma$ $=$ $2\mu$ $-$ $1$).

\centerline{\epsfxsize12.0cm\epsfbox[140 150  450 300]{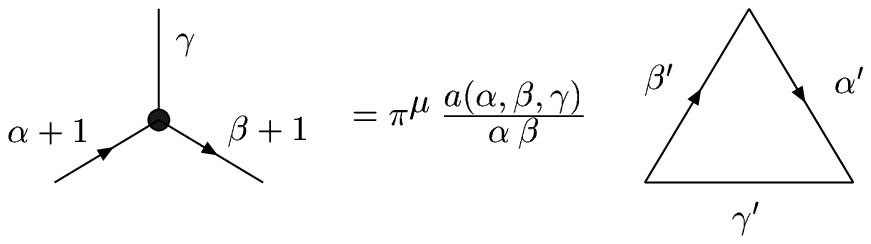}}
\vskip -1cm

\noindent
Vector-fermion vertex ($\alpha$ $+$ $\beta$ $+$ $\gamma$ $=$ $2\mu$ $-$ $1$).

\centerline{\epsfxsize12.0cm\epsfbox[140 150  500 300]{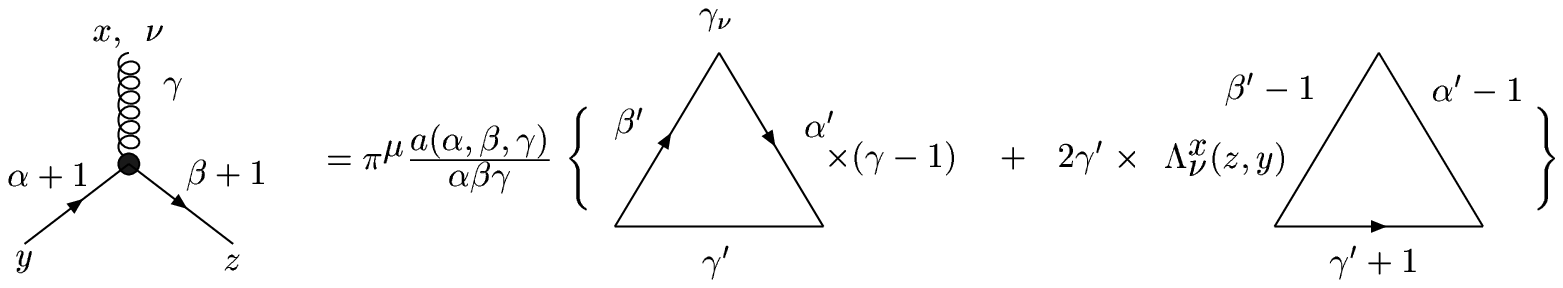}}
\vskip -1cm

\noindent
Here
\[ 
\Lambda^x_\nu(z,y) ~=~ \frac{(z-x)_\nu}{(z-x)^2}-\frac{(y-x)_\nu}{(y-x)^2} ~.
\]
As usual the line with an arrow with index $\alpha$ denotes the fermion 
propagator, $\xslash/(x^2)^\alpha$, a simple line corresponds to the boson 
propagator $1/(x^2)^\alpha$ and a wavy one is used to denote the conformal 
propagator~(\ref{confprop}).

\sect{Renormalization constants.} 
In this appendix we collect the values of the renormalization constants $Z_i$ 
and the basic RG functions at $O(1/\Nf)$. The calculations are rather 
straightforward and so we only list the results. We found 
\begin{eqnarray}
Z_{1}&=& 1 ~-~ \frac{g C_F \eta_{0}}{2n\Delta} \left( 
1 ~+~ \frac{\xi\mu}{(2\mu-1)(\mu-2)}\right) \nonumber \\
Z_{2}&=& 1 ~-~ \frac{g\eta_{0}}{2n\Delta}\left(C_{F}-C_A/2\right)
\left( 1 ~+~ \frac{\xi\mu}{(2\mu-1)(\mu-2)}\right) ~-~ 
\frac{g^2 C_A \eta_0}{8n\Delta} \left( \frac{\mu-1+\xi}{\mu-2} \right) 
\nonumber \\
Z_{3}&=& 1 ~+~ \frac{g\lambda^2 C_A \eta_{0}}{4n\Delta(\mu-2)} \left( 
1 ~-~ \frac{\xi}{(2\mu-1)}\right) \nonumber \\
Z_{4}&=& 1 ~-~ \frac{g \xi \lambda C_A \eta_{0}}{8n(\mu-2)\Delta} 
\left( \frac{\lambda}{(2\mu-1)} ~+~ \frac{g}{2} \right) 
\end{eqnarray}
where
\begin{equation}
\eta_{0} ~=~ \frac{(\mu-2)(2\mu-1)\Gamma(2\mu)} 
{\Gamma^2(\mu)\Gamma(\mu+1)\Gamma(2-\mu)} 
\end{equation}
and $g$ is defined in (\ref{mG}). Using (\ref{sf}), (\ref{so}) and 
(\ref{explicitRG}) one finds
\begin{eqnarray}
\label{indices}
\eta^{(1)}&=&2\gamma_\psi^{(1)} ~=~ C_F \eta_{0} \left( 
1 ~+~ \frac{\xi\mu}{(2\mu-1)(\mu-2)}\right)  \\
\gamma_c^{(1)}&=& -~ \frac{C_A\eta_{0}}{4(\mu-2)}
\left( 1 ~-~ \frac{\xi}{(2\mu-1)}\right) \\
\gamma_A^{(1)}&=&\frac{C_A\eta_0}{2(\mu-2)}
\left(1 ~+~ \frac{\xi(\mu-1)}{(2\mu-1)}\right)
\end{eqnarray}
For $\beta_\lambda$ we obtain
\begin{equation}
\beta_\lambda ~=~ \lambda(\gamma_A+2\gamma_c-\gamma_4) ~=~ 
\frac{\lambda (\lambda-1) C_A \eta_0}{4(\mu-2)(2\mu-1)} \Biggl[ 
\lambda(\xi-2(2\mu-1))-2(\mu-1)\Biggl(\xi+\frac{(2\mu-1)}{(\mu-1)} \Biggr)
\Biggr] ~.
\end{equation}
One can see that $\lambda$~$=$~$1$ is indeed a zero of the beta function. Also,
for $\xi_A$ we find
\begin{equation}
\xi_A^{(1)} ~=~ -~ \frac{(2\mu-1)}{(\mu-1)} ~
\end{equation} 
and $\beta_{\lambda}$ is simplified in this gauge to  
\begin{equation}
\beta_\lambda ~=~ -~ \frac{(2\mu-1) C_A \eta_0}{4(\mu-2)(\mu-1)}\lambda^2 
(\lambda-1) ~.
\end{equation}

Finally, for completeness we record the value of the amplitude combination 
$\tilde{z}$ required in the SD formalism at $O(1/\Nf^2)$. It is 
\begin{equation} 
\tilde{z}_1 ~=~ \frac{\Gamma(\mu+1)\eta_0}{2(2\mu-1)(\mu-2)} 
\end{equation} 
and 
\begin{eqnarray} 
\tilde{z}_2 &=& \frac{3\mu(\mu-1)\Gamma(\mu+1) C_F \eta_0^2} 
{2(2\mu-1)^2(\mu-2)^2} \left[ \Theta(\mu) ~-~ \frac{1}{(\mu-1)^2} \right]  
\nonumber \\ 
&& +~ \frac{\mu(\mu-1)\Gamma(\mu+1) C_A \eta_0^2}{4(2\mu-1)^2(\mu-2)^2} \left[ 
3\Theta(\mu) ~+~ \frac{(8\mu^5-92\mu^4+270\mu^3-301\mu^2+124\mu-12) 
\Psi(\mu)}{2\mu(\mu-1)(2\mu-1)(2\mu-3)(\mu-2)} \right. \nonumber \\ 
&& \left. ~~~~~~~~~~~-~ \Psi^2(\mu) ~-~ \Phi(\mu) ~+~ 
\frac{\bar{\xi}^2}{4(\mu-1)^2} ~+~ \frac{(\mu^2+2\mu-2)\bar{\xi}} 
{2(\mu-1)^2\mu^2} \right. \nonumber \\
&& \left. ~~~~~~~~~~~-~ \frac{(16\mu^7-120\mu^6+420\mu^5-776\mu^4+742\mu^3 
-349\mu^2+84\mu-12)}{2(2\mu-1)(2\mu-3)(\mu-1)^2(\mu-2)\mu^2} \right]  
\end{eqnarray} 
where the gluon propagator in the SD formalism is taken as  
\begin{equation} 
\tilde{A}_{\mu\nu}(k) ~=~ \frac{\tilde{B}}{(k^2)^{\mu-\beta}} \left[ 
\eta_{\mu\nu} ~-~ (1-\bar{\xi})\frac{k_\mu k_\nu}{k^2} \right] 
\end{equation} 
in momentum space. These will be useful in the calculation of gauge independent
critical exponents at $O(1/\Nf^2)$. For example, if one chooses a particular 
gauge to simplify those calculations, such as the Feynman gauge used in 
standard perturbative calculations, then one will require the value of the 
relevant variables for a general gauge parameter.

\end{document}